\pdfoutput=1
\documentclass[article,a4paper,11pt,fleqn,oneside]{memoir} 
\counterwithout{section}{chapter} 

\usepackage{Huhh_RC1} 
\usepackage{inputenc}

\usepackage{hyperref}
\usepackage{memhfixc}
\usepackage{longtable} 
\usepackage{array} 
\usepackage[clockwise]{rotating} 
\usepackage{threeparttable} 
\newsubfloat{figure} 

\numberwithin{corollary}{proposition} 

\newcommand\inB[2]{$\underset{\tiny{#2}}{\textbf{\em #1}}$}
\newcommand\tunder[2]{{#1}_{_{#2}}}

\author{%
Huhh, Jun-Sok\thanks{Department of Economics, Seoul National University. Tel: 82-2-10-4932-9881,  E-mail: \href{mailto:anarinsk@gmail.com}{anarinsk@gmail.com}
}%
}%

\title{%
{\Large\bfseries The Role of Opportunistic Punishment \\[-0.4cm] in the Evolution of Cooperation:} \\[-0.1cm]%
{\normalsize\bfseries An application of stochastic dynamics to public good game}\thanks{This version is made only for approved reviewers by the author. If you would refer or have any comment on this article, please contact author via below E-mail.}%
}

\def\CDLonesix#1{
\begin{tikzpicture}[scale=#1,
    knoten/.style={
      shade=ball,
      circle,
      inner sep=.1cm,
      circular drop shadow,
      draw}
    ]
  
 \node at (1,1) (CP) [knoten, ball color=green!70!black] {\inB{C}{39.0{\%}}};  
 \node at (6,1) (DP) [knoten, ball color=green!70!black] {\inB{D}{12.9{\%}}};
 \node at (6,6) (LP)  [knoten, ball color=green!70!black] {\inB{L}{48.1{\%}}};
  
 \draw [-stealth,line width=1.5] (CP) -- (DP);
 \draw [-stealth,line width=1.5] (DP) -- (LP);
 \draw [-stealth,line width=1.5] (LP) -- (CP);
 
  \node at (3.5,1.3) (CDF) [rotate=0]     {\small 0.07};  
  \node at (6.3,3.5) (DLF) [rotate=-90] {\small 0.08};  
  \node at (3.0,3.5) (DLF) [rotate=45] {\small 0.25};  
  \end{tikzpicture}
}

\def\CDPonesix#1{
\begin{tikzpicture}[scale=#1,
    knoten/.style={
      shade=ball,
      circle,
      inner sep=.1cm,
      circular drop shadow,
      draw}  
    ]
  
 \node at (1,1) (CP) [knoten, ball color=green!70!black] {\inB{C}{~~0{\%}~~}};  
 \node at (6,1) (DP) [knoten, ball color=green!70!black] {\inB{D}{100{\%}}};
 \node at (1,6) (PP)  [knoten, ball color=purple!70!white] {\inB{P}{~~0{\%}~~}};
  
 \draw [-stealth,line width=1.5] (CP) -- (DP);
 \draw [-,line width=1.5,style=dashed] (CP) -- (PP);
 
  \node at (3.5,1.3) (CDF) [rotate=0]     {\small 0.08};  
   \end{tikzpicture}
}

\def\CDLPonesix#1{
\begin{tikzpicture}[scale=#1,
    knoten/.style={
      shade=ball,
      circle,
      inner sep=.1cm,
      circular drop shadow,
      draw}  
    ]
  
 \node at (1,1) (CP) [knoten, ball color=green!70!black] {\inB{C}{14.1{\%}}};  
 \node at (6,1) (DP) [knoten, ball color=green!70!black] {\inB{D}{~4.7{\%}~}};
 \node at (6,6) (LP)  [knoten, ball color=green!70!black] {\inB{L}{~8.7{\%}~}};
 \node at (1,6) (PP)  [knoten, ball color=purple!70!white] {\inB{P}{72.5{\%}}};
  
 \draw [-stealth,line width=1.5] (CP) -- (DP);
 \draw [-stealth,line width=1.5] (DP) -- (LP);
 \draw [-stealth,line width=1.5] (LP) -- (CP);
 \draw [-stealth,line width=1.5] (LP) -- (PP);
 \draw [-,line width=1.5,style=dashed] (PP) -- (CP);
 
  \node at (3.5,1.3) (CtoP) [rotate=0] {\small 0.08};  
  \node at (6.3,3.5) (DtoL) [rotate=-90] {\small 0.25};  
  \node at (3.0,3.5) (LtoC) [rotate=45] {\small 0.07};  
  \node at (3.5,6.3) (LtoP) [rotate=0] {\small 0.07};  
\end{tikzpicture}
}

\def\lowbetaonesix#1{
\begin{tikzpicture}[scale=#1,
    knoten/.style={
      shade=ball,
      circle,
      inner sep=.1cm,
      circular drop shadow,
      draw}  
    ]
  
 \node at (1,1) (CP) [knoten, ball color=green!70!black] {\inB{C}{19.8{\%}}};  
 \node at (6,1) (DP) [knoten, ball color=green!70!black] {\inB{D}{10.7{\%}}};
 \node at (6,6) (LP)  [knoten, ball color=green!70!black] {\inB{L}{19.9{\%}}};
 \node at (1,6) (PP)  [knoten, ball color=purple!70!white] {\inB{P}{49.6{\%}}};
  
 \draw [-stealth,line width=1.5] (CP) -- (DP);
 \draw [-stealth,line width=1.5] (DP) -- (LP);
 \draw [-stealth,line width=1.5] (LP) -- (CP);
 \draw [-stealth,line width=1.5] (LP) -- (PP);
 \draw [-,line width=1.5,style=dashed] (PP) -- (CP);
 \draw [-stealth,line width=1.5] (PP) -- (DP);
 
 \node at (2.5,3) (LtoC) [rotate=45] {\small 0.07};  
 \node at (3.5,1.3) (CtoD) [rotate=0] {\small 0.08};  
 \node at (6.3,3.5) (DtoL) [rotate=-90] {\small 0.25};  
 \node at (3.5,6.3) (LtoP) [rotate=0] {\small 0.07};  
 \node at (4.5,3) (PtoD) [rotate=-45] {\small 0.02};  
\end{tikzpicture}
}

\def\lowsigmaonesix#1{
\begin{tikzpicture}[scale=#1,
    knoten/.style={
      shade=ball,
      circle,
      inner sep=.1cm,
      circular drop shadow,
      draw}  
    ]
  
 \node at (1,1) (CP) [knoten, ball color=green!70!black] {\inB{C}{11.0{\%}}};  
 \node at (6,1) (DP) [knoten, ball color=green!70!black] {\inB{D}{27.8{\%}}};
 \node at (6,6) (LP)  [knoten, ball color=green!70!black] {\inB{L}{~4.4{\%}~}};
 \node at (1,6) (PP)  [knoten, ball color=purple!70!white] {\inB{P}{56.8{\%}}};
  
 \draw [-stealth,line width=1.5] (CP) -- (DP);
 \draw [-stealth,line width=1.5] (DP) -- (LP);
 \draw [-stealth,line width=1.5] (LP) -- (CP);
 \draw [-stealth,line width=1.5] (LP) -- (PP);
 \draw [-,line width=1.5,style=dashed] (PP) -- (CP);
 
 \node at (2.5,3) (LtoC) [rotate=45] {\small 0.1};  
 \node at (3.5,1.3) (CtoD) [rotate=0] {\small 0.08};  
 \node at (6.3,3.5) (DtoL) [rotate=-90] {\small 0.03};  
 \node at (3.5,6.3) (LtoP) [rotate=0] {\small 0.1};  
\end{tikzpicture}
}

\def\lowgammalowsigmaonesix#1{
\begin{tikzpicture}[scale=#1,
    knoten/.style={
      shade=ball,
      circle,
      inner sep=.1cm,
      circular drop shadow,
      draw}  
    ]
  
 \node at (1,1) (CP) [knoten, ball color=green!70!black] {\inB{C}{11.8{\%}}};  
 \node at (6,1) (DP) [knoten, ball color=green!70!black] {\inB{D}{46.8{\%}}};
 \node at (6,6) (LP)  [knoten, ball color=green!70!black] {\inB{L}{~7.3{\%}~}};
 \node at (1,6) (PP)  [knoten, ball color=purple!70!white] {\inB{P}{34.1{\%}}};
  
 \draw [-stealth,line width=1.5] (CP) -- (DP);
 \draw [-stealth,line width=1.5] (DP) -- (LP);
 \draw [-stealth,line width=1.5] (LP) -- (CP);
 \draw [-stealth,line width=1.5] (LP) -- (PP);
 \draw [-,line width=1.5,style=dashed] (PP) -- (CP);
 \draw [-stealth,line width=1.5] (PP) -- (DP);
 
 \node at (2.5,3) (LtoC) [rotate=45] {\small 0.1};  
 \node at (3.5,1.3) (CtoD) [rotate=0] {\small 0.08};  
 \node at (6.3,3.5) (DtoL) [rotate=-90] {\small 0.03};  
 \node at (3.5,6.3) (LtoP) [rotate=0] {\small 0.1};  
 \node at (4.5,3) (PtoD) [rotate=-45] {\small 0.02};  
\end{tikzpicture}
}

\def\highbetahighsigmasevensix#1{
\begin{tikzpicture}[scale=#1,
    knoten/.style={
      shade=ball,
      circle,
      inner sep=.1cm,
      circular drop shadow,
      draw}  
    ]
  
 \node at (1,1) (CP) [knoten, ball color=green!70!black] {\inB{C}{44.2{\%}}};  
 \node at (6,1) (DP) [knoten, ball color=green!70!black] {\inB{D}{14.6{\%}}};
 \node at (6,6) (LP)  [knoten, ball color=green!70!black] {\inB{L}{27.2{\%}}};
 \node at (1,6) (PP)  [knoten, ball color=purple!70!white] {\inB{P}{14.0{\%}}};
  
 \draw [-stealth,line width=1.5] (CP) -- (DP);
 \draw [-stealth,line width=1.5] (DP) -- (LP);
 \draw [-stealth,line width=1.5] (LP) -- (CP);
 \draw [-stealth,line width=1.5] (LP) -- (PP);
 \draw [-stealth,line width=1.5] (PP) -- (CP);
 
 \node at (2.5,3) (LtoC) [rotate=45] {\small 0.07};  
 \node at (3.5,1.3) (CtoD) [rotate=0] {\small 0.08};  
 \node at (6.3,3.5) (DtoL) [rotate=-90] {\small 0.25};  
 \node at (3.5,6.3) (LtoP) [rotate=0] {\small 0.07};  
 \node at (0.7,3.5) (PtoC) [rotate=90] {\small 0.13};  
\end{tikzpicture}
}

\def\lowbetahighsigmasevensix#1{
\begin{tikzpicture}[scale=#1,
    knoten/.style={
      shade=ball,
      circle,
      inner sep=.1cm,
      circular drop shadow,
      draw}  
    ]
  
 \node at (1,1) (CP) [knoten, ball color=green!70!black] {\inB{C}{~3.4{\%}~}};  
 \node at (6,1) (DP) [knoten, ball color=green!70!black] {\inB{D}{~3.6{\%}~}};
 \node at (6,6) (LP)  [knoten, ball color=green!70!black] {\inB{L}{~6.8{\%}~}};
 \node at (1,6) (PP)  [knoten, ball color=purple!70!white] {\inB{P}{86.2{\%}}};
  
 \draw [-stealth,line width=1.5] (CP) -- (DP);
 \draw [-stealth,line width=1.5] (DP) -- (LP);
 \draw [-stealth,line width=1.5] (LP) -- (CP);
 \draw [-stealth,line width=1.5] (LP) -- (PP);
 \draw [stealth-,line width=1.5] (PP) -- (CP);
 
 \node at (2.5,3) (LtoC) [rotate=45] {\small 0.07};  
 \node at (3.5,1.3) (CtoD) [rotate=0] {\small 0.08};  
 \node at (6.3,3.5) (DtoL) [rotate=-90] {\small 0.25};  
 \node at (3.5,6.3) (LtoP) [rotate=0] {\small 0.07};  
 \node at (0.7,3.5) (PtoC) [rotate=90] {\small 0.05};  
\end{tikzpicture}
}

\def\highbetalowsigmasevensix#1{
\begin{tikzpicture}[scale=#1,
    knoten/.style={
      shade=ball,
      circle,
      inner sep=.1cm,
      circular drop shadow,
      draw}  
    ]
  
 \node at (1,1) (CP) [knoten, ball color=green!70!black] {\inB{C}{23.6{\%}}};  
 \node at (6,1) (DP) [knoten, ball color=green!70!black] {\inB{D}{59.8{\%}}};
 \node at (6,6) (LP)  [knoten, ball color=green!70!black] {\inB{L}{~9.3{\%}~}};
 \node at (1,6) (PP)  [knoten, ball color=purple!70!white] {\inB{P}{~7.3{\%}~}};
  
 \draw [-stealth,line width=1.5] (CP) -- (DP);
 \draw [-stealth,line width=1.5] (DP) -- (LP);
 \draw [-stealth,line width=1.5] (LP) -- (CP);
 \draw [-stealth,line width=1.5] (LP) -- (PP);
 \draw [-stealth,line width=1.5] (PP) -- (CP);
 
 \node at (2.5,3) (LtoC) [rotate=45] {\small 0.1};  
 \node at (3.5,1.3) (CtoD) [rotate=0] {\small 0.08};  
 \node at (6.3,3.5) (DtoL) [rotate=-90] {\small 0.03};  
 \node at (3.5,6.3) (LtoP) [rotate=0] {\small 0.1};  
 \node at (0.7,3.5) (PtoC) [rotate=90] {\small 0.13};  
\end{tikzpicture}
}

\def\lowbetalowgammalowsigmasevensix#1{
\begin{tikzpicture}[scale=#1,
    knoten/.style={
      shade=ball,
      circle,
      inner sep=.1cm,
      circular drop shadow,
      draw}  
    ]
  
 \node at (1,1) (CP) [knoten, ball color=green!70!black] {\inB{C}{~0.8{\%}~}};  
 \node at (6,1) (DP) [knoten, ball color=green!70!black] {\inB{D}{~6.8{\%}~}};
 \node at (6,6) (LP)  [knoten, ball color=green!70!black] {\inB{L}{~1.1{\%}~}};
 \node at (1,6) (PP)  [knoten, ball color=purple!70!white] {\inB{P}{91.3{\%}}};
  
 \draw [-stealth,line width=1.5] (CP) -- (DP);
 \draw [-stealth,line width=1.5] (DP) -- (LP);
 \draw [-stealth,line width=1.5] (LP) -- (CP);
 \draw [-stealth,line width=1.5] (LP) -- (PP);
 \draw [stealth-,line width=1.5] (PP) -- (CP);
 
 \node at (2.5,3) (LtoC) [rotate=45] {\small 0.1};  
 \node at (3.5,1.3) (CtoD) [rotate=0] {\small 0.08};  
 \node at (6.3,3.5) (DtoL) [rotate=-90] {\small 0.03};  
 \node at (3.5,6.3) (LtoP) [rotate=0] {\small 0.01};  
 \node at (0.7,3.5) (PtoC) [rotate=90] {\small 0.06};  
\end{tikzpicture}
}

\def\deltawork#1{
\begin{tikzpicture}[scale=#1]
    \begin{axis}[
        xlabel=$\delta$,
        ylabel=frequency of cooperation,
       legend style={at={(0.3,0.95)}}]
    \addplot[smooth,mark=*,blue] plot coordinates {
 (0.5, 0.)
 (0.51, 0.)
 (0.52, 0.)
 (0.53, 0.)
 (0.54, 0.)
 (0.55, 0.)
 (0.56, 0.)
 (0.57, 0.)
 (0.58, 0.)
 (0.59, 0.)
 (0.6, 0.)
 (0.61, 0.)
 (0.62, 0.)
 (0.63, 0.)
 (0.64, 0.)
 (0.65, 0.)
 (0.66, 0.)
 (0.67, 0.)
 (0.68, 20.3169)
 (0.69, 21.7061)
 (0.7, 23.162)
 (0.71, 24.6841)
 (0.72, 26.2714)
 (0.73, 27.9222)
 (0.74, 29.6343)
 (0.75, 31.4049)
 (0.76, 33.2304)
 (0.77, 35.1068)
 (0.78, 37.0293)
 (0.79, 38.9928)
 (0.8, 40.9914)
 (0.81, 43.019)
 (0.82, 45.0689)
 (0.83, 47.1342)
 (0.84, 49.2079)
 (0.85, 51.2826)
 (0.86, 53.3513)
 (0.87, 55.4068)
 (0.88, 57.4422)
 (0.89, 100.)
 (0.9, 100.)
 (0.91, 100.)
 (0.92, 100.)
 (0.93, 100.)
 (0.94, 100.)
 (0.95, 100.)
 (0.96, 100.)
 (0.97, 100.)
 (0.98, 100.)
 (0.99, 100.)
 (1., 100.)
    };
    \addlegendentry{{\tiny$CDP$}}

    \addplot[smooth,color=red,mark=x]
        plot coordinates {
 (0.5, 60.908)
 (0.51, 62.1489)
 (0.52, 63.3875)
 (0.53, 64.6215)
 (0.54, 65.8488)
 (0.55, 67.0673)
 (0.56, 68.2748)
 (0.57, 70.151)
 (0.58, 71.3353)
 (0.59, 72.4995)
 (0.6, 73.6421)
 (0.61, 74.7615)
 (0.62, 75.8563)
 (0.63, 76.9255)
 (0.64, 77.9677)
 (0.65, 78.9821)
 (0.66, 79.9678)
 (0.67, 80.9242)
 (0.68, 81.9893)
 (0.69, 82.876)
 (0.7, 83.7325)
 (0.71, 84.5586)
 (0.72, 85.3542)
 (0.73, 86.1195)
 (0.74, 86.8547)
 (0.75, 87.5601)
 (0.76, 88.236)
 (0.77, 88.8829)
 (0.78, 89.5012)
 (0.79, 90.0917)
 (0.8, 90.6549)
 (0.81, 91.1915)
 (0.82, 91.7022)
 (0.83, 92.1878)
 (0.84, 92.6492)
 (0.85, 93.087)
 (0.86, 93.5022)
 (0.87, 93.8955)
 (0.88, 94.2679)
 (0.89, 100.)
 (0.9, 100.)
 (0.91, 100.)
 (0.92, 100.)
 (0.93, 100.)
 (0.94, 100.)
 (0.95, 100.)
 (0.96, 100.)
 (0.97, 100.)
 (0.98, 100.)
 (0.99, 100.)
 (1., 100.)
        };
    \addlegendentry{{\tiny$CDLP$}}
    \end{axis}
    \end{tikzpicture}
}

\def\deltaerror#1{
\begin{tikzpicture}[scale=#1]
    \begin{axis}[
        xlabel=$\delta$,
        ylabel=frequency of cooperation,
       legend style={at={(0.3,0.95)}}]
    \addplot[smooth,mark=*,blue] plot coordinates {
  (0.5, 0.)
 (0.51, 0.)
 (0.52, 0.)
 (0.53, 0.)
 (0.54, 0.)
 (0.55, 0.)
 (0.56, 0.)
 (0.57, 0.)
 (0.58, 0.)
 (0.59, 0.)
 (0.6, 0.)
 (0.61, 0.)
 (0.62, 0.)
 (0.63, 0.)
 (0.64, 0.)
 (0.65, 0.)
 (0.66, 0.)
 (0.67, 0.)
 (0.68, 0.)
 (0.69, 0.)
 (0.7, 0.)
 (0.71, 0.)
 (0.72, 0.)
 (0.73, 0.)
 (0.74, 0.)
 (0.75, 0.)
 (0.76, 0.)
 (0.77, 0.)
 (0.78, 0.)
 (0.79, 0.)
 (0.8, 0.)
 (0.81, 0.)
 (0.82, 0.)
 (0.83, 0.)
 (0.84, 0.)
 (0.85, 0.)
 (0.86, 0.)
 (0.87, 0.)
 (0.88, 0.)
 (0.89, 0.)
 (0.9, 0.)
 (0.91, 0.)
 (0.92, 0.)
 (0.93, 0.)
 (0.94, 0.)
 (0.95, 0.)
 (0.96, 0.)
 (0.97, 0.)
 (0.98, 0.)
 (0.99, 0.)
 (1., 0.)
    };
    \addlegendentry{{\tiny$CDP$}}

    \addplot[smooth,color=red,mark=x]
        plot coordinates {
  (0.5, 27.8254)
 (0.51, 28.0015)
 (0.52, 28.1798)
 (0.53, 28.3605)
 (0.54, 28.5437)
 (0.55, 28.7293)
 (0.56, 28.9173)
 (0.57, 29.1079)
 (0.58, 29.3009)
 (0.59, 29.4966)
 (0.6, 29.6948)
 (0.61, 29.8957)
 (0.62, 30.0992)
 (0.63, 30.3055)
 (0.64, 30.5145)
 (0.65, 30.7262)
 (0.66, 30.9407)
 (0.67, 31.158)
 (0.68, 31.3782)
 (0.69, 31.6013)
 (0.7, 31.8273)
 (0.71, 32.0563)
 (0.72, 32.2882)
 (0.73, 32.5231)
 (0.74, 32.7611)
 (0.75, 33.0022)
 (0.76, 33.2463)
 (0.77, 33.4936)
 (0.78, 33.744)
 (0.79, 33.9976)
 (0.8, 34.2545)
 (0.81, 34.5145)
 (0.82, 34.7778)
 (0.83, 35.0444)
 (0.84, 35.3143)
 (0.85, 35.5875)
 (0.86, 35.8641)
 (0.87, 36.1441)
 (0.88, 36.4274)
 (0.89, 36.7141)
 (0.9, 37.0043)
 (0.91, 37.2979)
 (0.92, 37.595)
 (0.93, 37.8955)
 (0.94, 38.1996)
 (0.95, 38.5071)
 (0.96, 38.8181)
 (0.97, 39.1326)
 (0.98, 39.4506)
 (0.99, 39.7722)
 (1., 40.0973)
        };
    \addlegendentry{{\tiny$CDLP$}}
    \end{axis}
    \end{tikzpicture}
}
\def\CDPFermiCP#1{
\begin{tikzpicture}[scale=#1,
    knoten/.style={
      shade=ball,
      circle,
      inner sep=.1cm,
      circular drop shadow,
      draw},
      knotenn/.style={
      shade=ball,
      circle,
      inner sep=.23cm,
      circular drop shadow,
      draw}     
    ]
   
 \node at (1,1) (CP) [knotenn, ball color=green!70!black] {\Large $C$};  
 \node at (6,1) (DP) [knotenn, ball color=green!70!black] {\Large $D$};
 \node at (1,6) (PP) [knotenn, ball color=purple!70!white] {\Large $P$};
   
 \draw [-stealth,line width=1.5] (PP) -- (DP);
 \draw [-,line width=1.5,style=dashed] (PP) -- (CP);
   
 \node at (0.5,3.5) (CtoP) [rotate=90] {\small $\rho_{\cdot}=1/M$};  
 \node at (3.8,3.7) (DtoP) [rotate=-45] {\small for $\beta < \tunder{\beta}{CP}$};    
 \end{tikzpicture}
}

\def\CDPFermiOP#1{
\begin{tikzpicture}[scale=#1,
    knoten/.style={
      shade=ball,
      circle,
      inner sep=.1cm,
      circular drop shadow,
      draw},
      knotenn/.style={
      shade=ball,
      circle,
      inner sep=.23cm,
      circular drop shadow,
      draw}     
    ]
   
 \node at (1,1) (CP) [knotenn, ball color=green!70!black] {\Large $C$};  
 \node at (6,1) (DP) [knotenn, ball color=green!70!black] {\Large $D$};
 \node at (1,6) (PP) [knotenn, ball color=purple!70!white] {\Large $P$};
   
 \draw [-,line width=1.5,color=black!30!white] (PP) -- (DP);
 \draw [stealth-,line width=1.5] (PP) -- (CP);
   
 \node at (0.5,3.5) (CtoP) [rotate=90] {\small for $\delta \geq \tunder{\delta}{OP}$};  
 \node at (3.8,3.7) (DtoP) [rotate=-45] {\small $\tunder{\rho}{\cdot}=0$};    
 \end{tikzpicture}
}

\def\CDAPGOOD#1{
\begin{tikzpicture}[scale=#1]

\node [inner sep=0pt,above right] {\includegraphics[scale=#1]{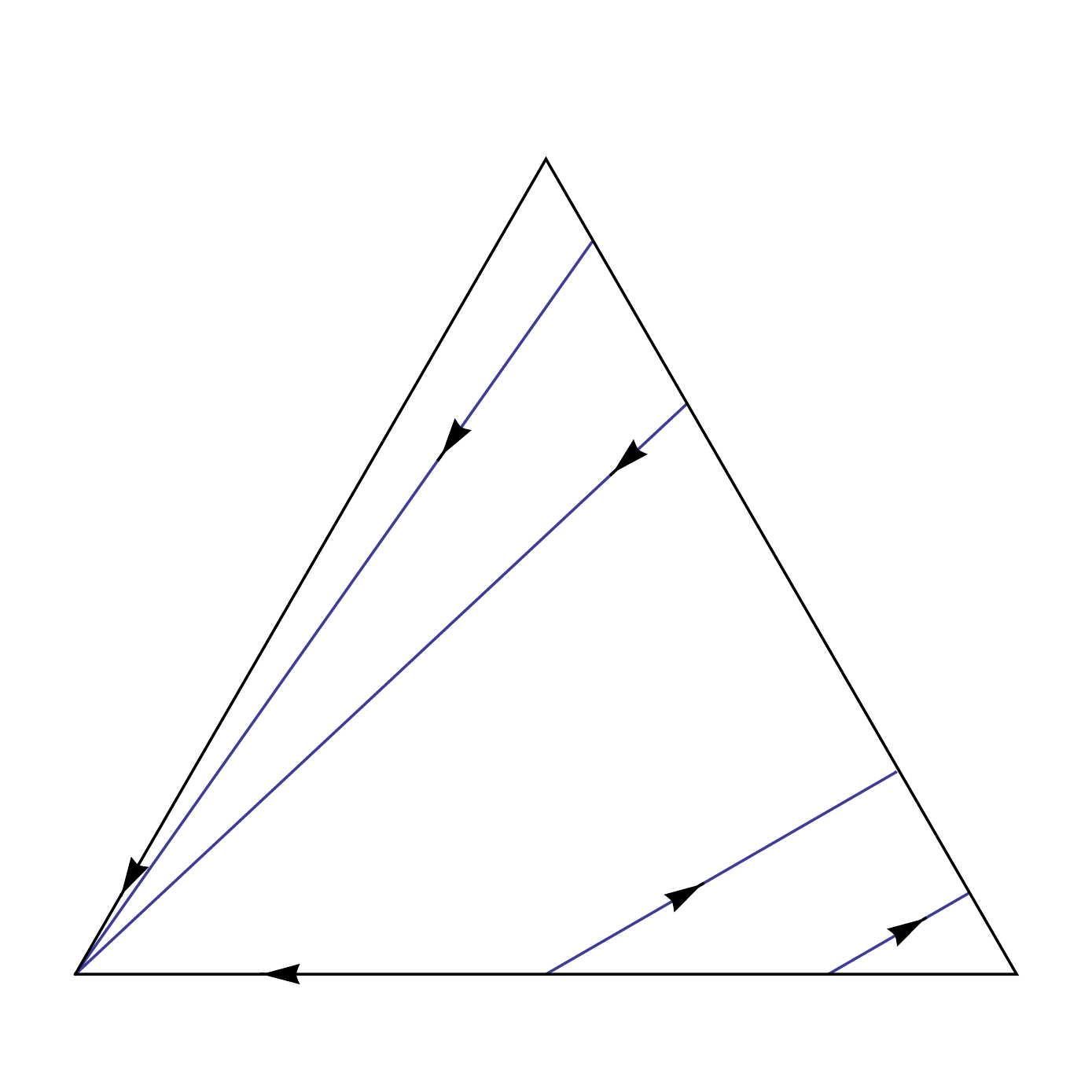}};

\draw[line width=0.5pt](10.6,5.9)--(6.1,1.5);
\draw[fill=white, line width=0.9pt] (6.1,1.5) circle (4pt);
\draw[fill=black, line width=0.9pt] (10.6,5.9) circle (4pt);
\draw[line width=1.2pt](10.6,5.9)--(13.1,1.5);

\draw[fill=white, line width=0.9pt] (7.05,12) circle (4pt);
\draw[fill=black, line width=0.9pt] (13.1,1.5) circle (4pt);
\draw[fill=black, line width=0.9pt] (1,1.5) circle (4pt);

\draw[->,>=stealth',line width=0.4pt,thick] (8.65,4) -- (8.64,3.99);  

\node at (7.05,13) {C};
\node at (14.1,1.2) {CP};
\node at (0,1.2) {D};

\end{tikzpicture}
}

\def\CDAPBAD#1{
\begin{tikzpicture}[scale=#1]

\node [inner sep=0pt,above right] 
                {\includegraphics[scale=#1]{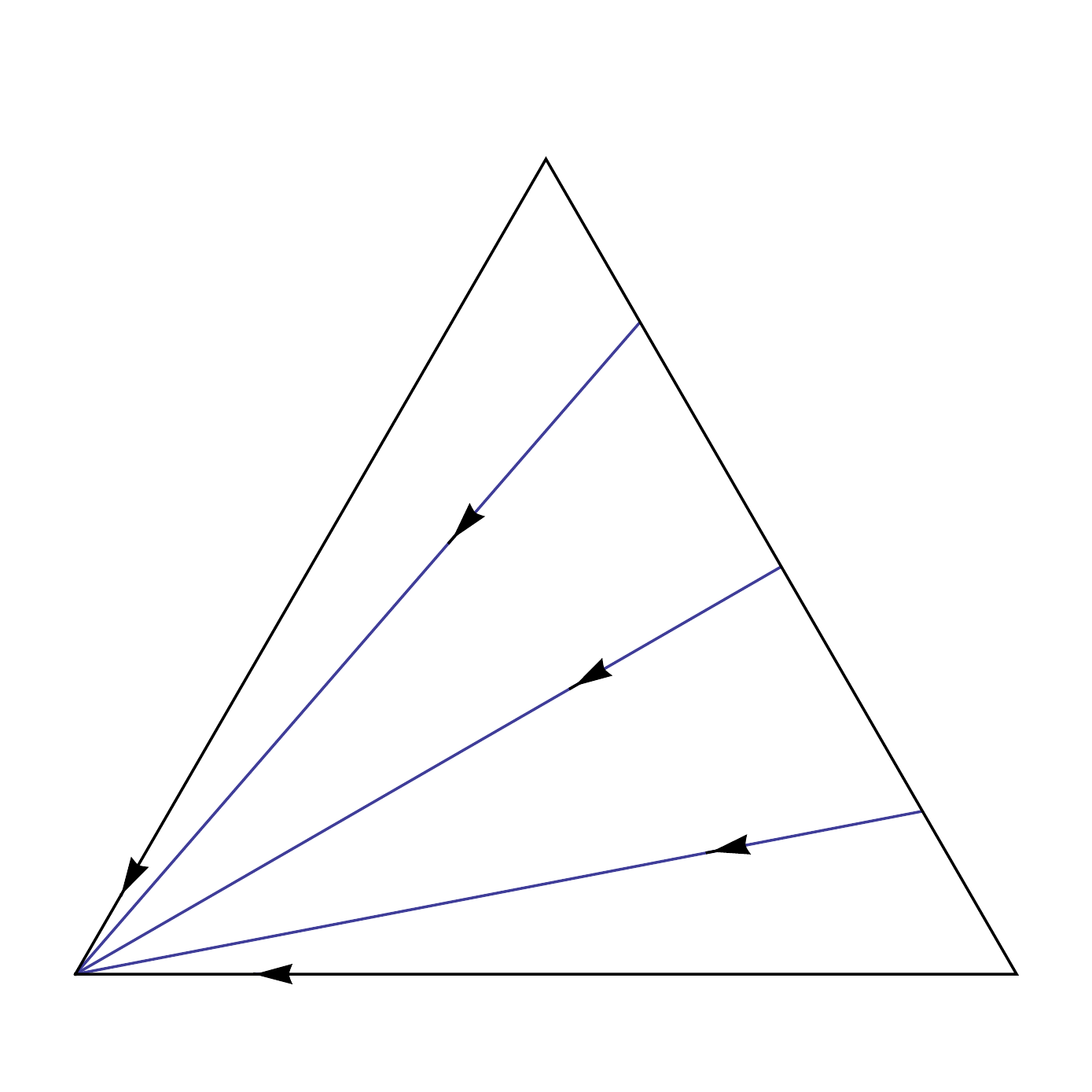}};

\draw[fill=white, line width=0.9pt] (7.05,12) circle (4pt);
\draw[fill=white, line width=0.9pt] (13.1,1.5) circle (4pt);
\draw[fill=black, line width=0.9pt] (1,1.5) circle (4pt);

\node at (7.05,13) {C};
\node at (14.1,1.2) {CP};
\node at (0,1.2) {D};

\end{tikzpicture}
}

\def\CDLCYCLE#1{
\begin{tikzpicture}[scale=#1]

\node [inner sep=0pt,above right] {\includegraphics[scale=#1]{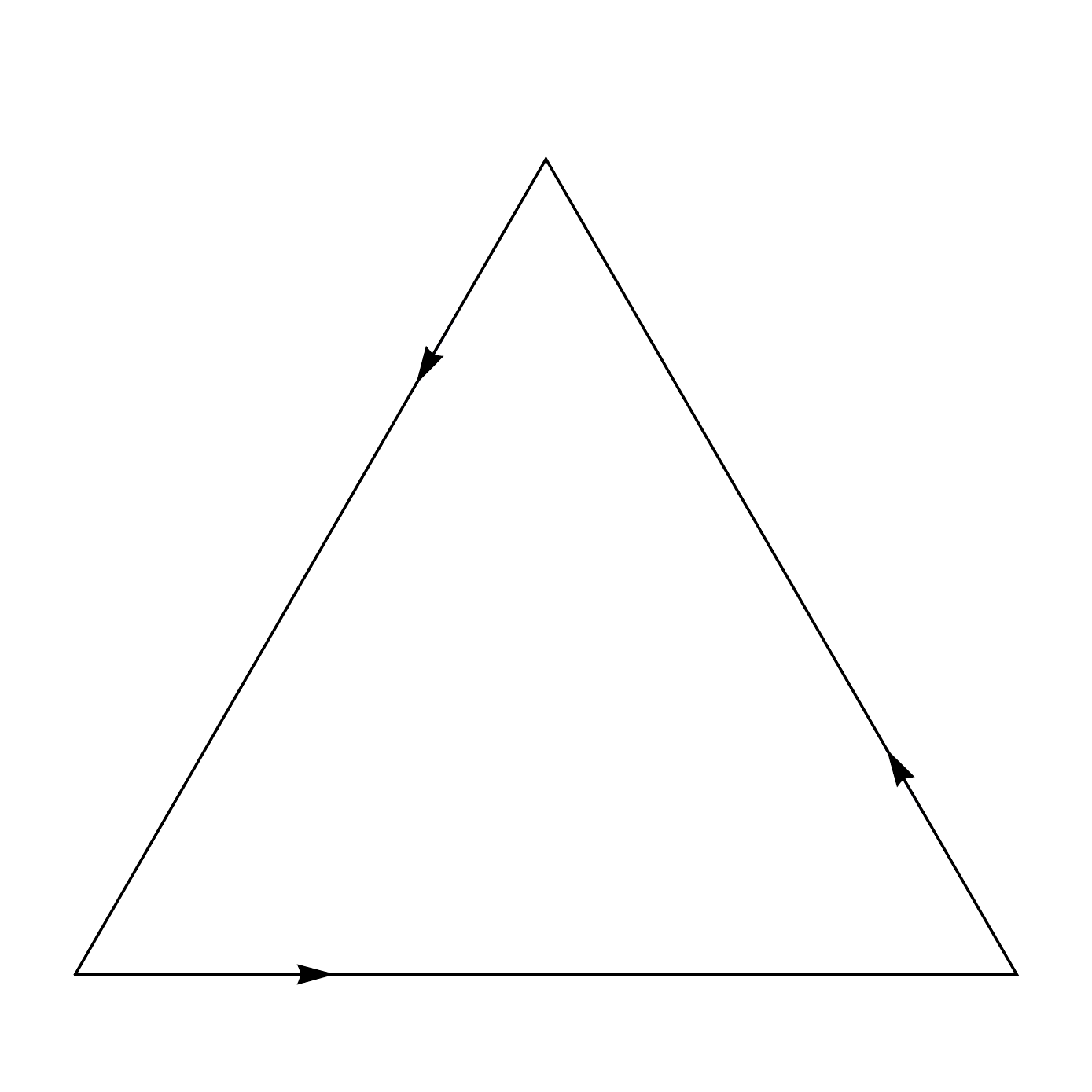}};

\draw[line width=0.5pt, color=blue!40!black] (7.05,5) circle (40pt);
\draw[fill=white,line width=0.5pt] (7.05,5) circle (4pt);
\draw[->,>=stealth',line width=0.4pt,thick] (8.45,5) -- (8.45,5.01);   

\draw[fill=white, line width=0.9pt] (7.05,12) circle (4pt);
\draw[fill=white, line width=0.9pt] (13.1,1.5) circle (4pt);
\draw[fill=white, line width=0.9pt] (1,1.5) circle (4pt);

\node at (7.05,13) {C};
\node at (14.1,1.2) {L};
\node at (0,1.2) {D};

\end{tikzpicture}
}

\def\CDOPGOOD#1{
\begin{tikzpicture}[scale=#1]

\node [inner sep=0pt,above right] 
                {\includegraphics[scale=#1]{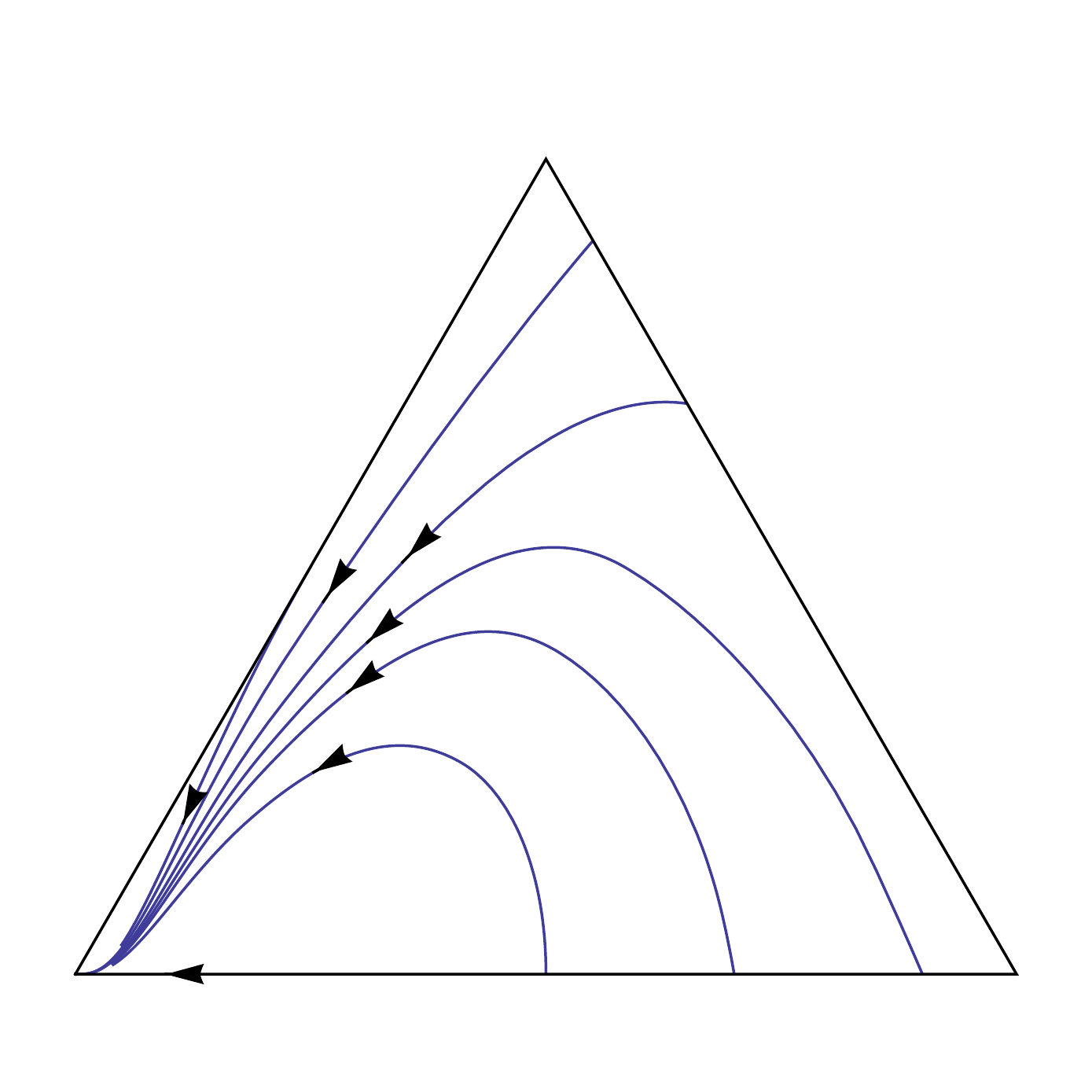}};
 
\draw[fill=white, line width=0.9pt] (7.05,12) circle (4pt);
\draw[fill=white, line width=0.9pt] (13.1,1.5) circle (4pt);
\draw[fill=black, line width=0.9pt] (1,1.5) circle (4pt);

\node at (7.05,13) {C};
\node at (14.1,1.2) {OP};
\node at (0,1.2) {D};

\end{tikzpicture}
}

\def\CDOPBAD#1{
\begin{tikzpicture}[scale=#1]

\node [inner sep=0pt,above right] 
                {\includegraphics[scale=#1]{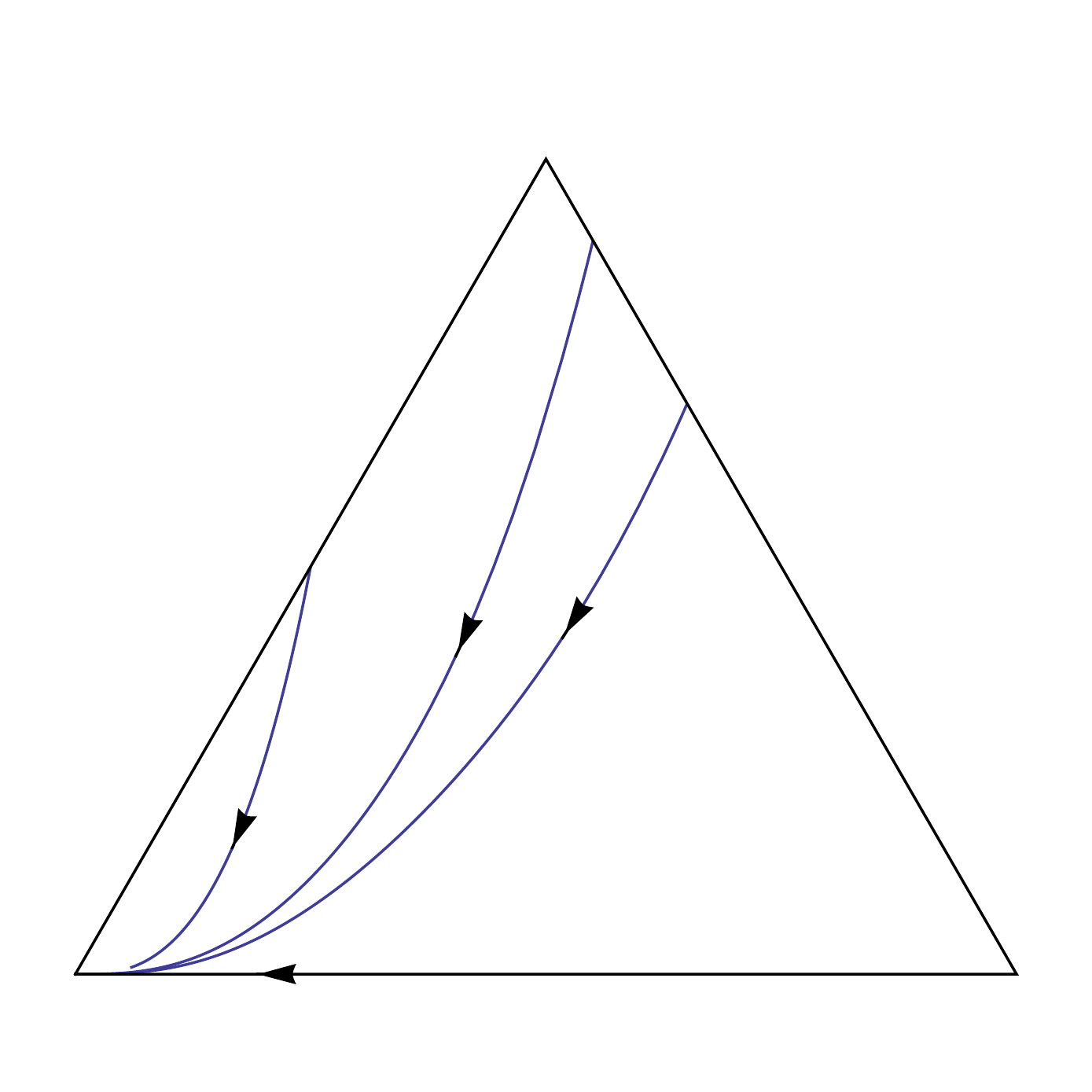}};

\draw[line width=0.5pt, color=blue!40!black] (9.5,3.5) circle (40pt);
\draw[fill=white,line width=0.5pt] (9.5,3.5) circle (4pt);

\draw[->,>=stealth',line width=0.4pt,thick] (7,1.52) -- (7.01,1.52);     
\draw[->,>=stealth',line width=0.4pt,thick] (11.15,5) -- (11.145,5.01);    
\draw[->,>=stealth',line width=0.4pt,thick] (7,4) -- (6.99,3.99);   
\draw[->,>=stealth',line width=0.4pt,thick] (10.9,3.6) -- (10.9,3.601);   

\draw[line width=0.5pt] (10,7)--(4.5,1.5);

\draw[fill=white, line width=0.9pt] (7.05,12) circle (4pt);
\draw[fill=black, line width=0.9pt] (10,7) circle (4pt);
\draw[fill=white, line width=0.9pt] (13.1,1.5) circle (4pt);
\draw[fill=black, line width=0.9pt] (1,1.5) circle (4pt);
\draw[fill=white, line width=0.9pt] (4.5,1.5) circle (4pt);

\node at (11,6.8) {\footnotesize$\tunder{p}{OP}$};
\node at (7.05,13) {C};
\node at (14.1,1.2) {OP};
\node at (0,1.2) {D};

\end{tikzpicture}
}

\begin{document}

\maketitle

{\linespread{1.3}
\begin{abstract} 
This paper discusses the role of opportunistic punisher who may act selfishly to free-ride cooperators or not to be exploited by defectors. To consider opportunistic punisher, we make a change to the sequence of one-shot public good game; instead of putting action choice first before punishment, the commitment of punishment is declared first before choosing the action of each participant. In this commitment-first setting, punisher may use information about her team, and may defect to increase her fitness in the team. Reversing sequence of public good game can induce different behavior of punisher, which cannot be considered in standard setting where punisher always chooses cooperation. Based on stochastic dynamics developed by evolutionary economists and biologists, we show that opportunistic punisher can make cooperation evolve where cooperative punisher fails. This alternative route for the evolution of cooperation relies paradoxically on the players' selfishness to profit from others' unconditional cooperation and defection. 
\noindent \keyword{public good game, stochastic (adaptive) dynamics, punishment, fixation probability} 
\end{abstract} 
}

\section{Introduction} 

Public good game (PGG) is one of the most active research themes in economics and evolutionary biology for last ten years. In typical PGG experiments, $G$ individuals have the opportunity to cooperate and bestow a fixed amount $q$ into a common resource, or to defect and input nothing. The total amount in the common is multiplied by a factor $r$ and equally distributed among the members without regard to their contributions. Average return of unit investment is $q r/G$ where $G$ is the size of a team. If $r<G$, a rational player does not contribute, and Nash equilibrium is universal defection. But, assuming that all of members cooperates, the return is $(r-1)q$ that is larger than that from defection. 

This is a classic social dilemma that the Nash equilibrium is different from the social optimum. As for the case of Prisoner's Dilemma, academic interests around PGG also have focused on ways and mechanisms that make the evolution of cooperation possible. For this purpose, besides from introducing reputation effect by repeating games, two theoretical methods has been proposed: 
\begin{inparaenum}[i)]
\item stern punishment that is unrelated to payoff consideration 
\item the option that players exit from the game. 
\end{inparaenum}

For the first direction, \citet{Fehr_Gachter:2000} shows experimental evidences by conducting two-stage version of PGG.\footnote{In stage one, four subjects play a simple PGG, and in stage two, contributions of individual members are revealed and any member of the four-player group may choose to reduce the earnings of any of the other members of the group at cost to himself.} Obviously, sub-game perfect equilibrium is that agents never punish in stage two as it lowers their payoff; hence punishment is not a factor in decisions in stage one; hence no contributions in stage one as usual. But, ample of experimental studies consistently show that availability of the punishment mechanism increases contributions markedly relative to their absence. Further, some punishment does occur actually. 

This influential work has been followed by numerous studies that explore around punishment, and this implies that punishment in PGG is a key part of institutional and behavioral mechanism to overcome the social dilemma of cooperation \citep{Ledyard:1997,Sigmund:2007}. But, a big piece of puzzle about punishment is that punishment itself cannot be evolutionarily favored because this behavior cannot get higher payoff or fitness than defection. Let us imagine the situation that all of population consists of defectors. When a mutant punisher comes about, her payoff cannot exceed that of other defectors as long as sufficient cost of punishment is imposed. Even though the role of punishment in the evolution of cooperation in PGG may be reasonably accepted, this behavior may not be selected and survive in evolutionary process. 

Another direction of research based on evolutionary dynamics tries the power of exit options that makes players avoid worst outcomes in PGG. The idea is that defection in PGG can be circumvented by making players choose an option that has an intermediate value between universal defection and high-frequency cooperation. \citet{Brandt_Hauert_Sigmund:2003} shows that exit option makes evolutionary cycle among cooperation, defection and exit by replicator dynamics. According to its conclusion, however, this evolutionary cycle by exit option cannot help the evolution of cooperation in that three strategy enjoys same payoff or fitness, which means that the participation in PGG is not better than exit option.

\citet{Hauert_Traulsen_Brandt_Nowak_Sigmund:2007} proposes that two directions may be interwoven in the evolution of cooperation. Their Intuition is that defectors may break the homogeneous population of cooperators, but the equilibrium based on universal defection can be also shaken by the exit option. When all of population chooses exiting, cooperation or cooperation with punishment is better choice for players. If we focus on homogeneous sates, these four states are in evolutionary cycles. Based on evolutionary dynamics, \citet{Hauert_Traulsen_Brandt_Nowak_Sigmund:2007} shows that cooperative state can be dominant state, which is a route for the evolution of cooperation in one-shot PGG game. 

Based on former studies, this paper tries to consider an unexplored theoretical element in PGG by stochastic evolutionary dynamics. We investigate the role and the effectiveness of punishment in PGG by assuming slightly different setting of PGG and behavioral pattern of punisher. For this, the sequence of standard PGG, strategy choice first and punishment with the information of players' action, is reversely arranged. So to speak, players commit their punishment first, and choices action with this information. Also, we introduce the opportunistic punisher who chooses its strategy based on the number of punisher in her team. Hence, for this new type of punisher, the information about punishing commitment plays a key role in choosing their actions. Even though defection of the punisher can hurt herself if she chooses defection, this choice may pay when there is sufficiently large number of cooperators. Intuitively, different from \citet{Hauert_Traulsen_Brandt_Nowak_Sigmund:2007} where the punisher is originated from cooperator, our punisher is opportunistic in that they deviates from cooperative strategy when it pays.

The organization of the paper is following: 
\Sref{SEC:Method} succinctly describes the basic of our PGG and methodology we reply on, stochastic (adaptive) dynamics. \Sref{SEC:Numerical} shows numerical cases our main discussion. \Sref{SEC:Fermi} presents theoretical extension of this paper with assuming \Sref{SEC:Concluding} is concluding remarks.

\section {Setup and Method}\label{SEC:Method}
\subsection{PGG, Nash equilibrium and Punishment}

This paper is based on a $G$-person game called Public Good Game. We consider a well-mixed population of constant size $M \geq 2$, and $G$ individuals are randomly selected and offered the option to participate PGG. Each should decide whether to contribute for the public good or not; cooperate ($C$) or defect ($D$). For simplicity, players invest fixed an amount $c$, we assume that the contributions of all $\tunder{G}{C}$ cooperators are multiplied by $r>1$ and then divided among all $G$ players participating in the game. The payoff for each $C$ and $D$ are given by 

\begin{align*}
\begin{cases}
\frac{r c\, \tunder{G}{C}}{G}~&\text{for $D$} \\
\frac{r c\, \tunder{G}{C}}{G}-c~&\text{for $C$}.
\end{cases}
\end{align*} 

Nash equilibrium is easily given by considering the benefit generated by switching from $C$ to $D$, which is $c(1-\frac{r}{G}).$ It is obvious that players would play $D$ as long as $r<G$. This is the social dilemma that resembles Prisoner's Dilemma. Most of interests around evolutionary game theory lie in finding routes or mechanism to overcome this uncooperative state. Can this social dilemma be evaded through positive or negative devices specifically directed towards individual players? In this paper, we shall focus on negative and neutral mechanism: punishment and exit.\footnote{Recently, some experimental evidences show that positive devices can be more effective in inducing cooperation among participants. In this paper, we remain around punishment issue, which has been more intensively discussed topics.}

When she quits, $\sigma$ is her payoff. We call her the Loner ($L$). When she participates, her types differentiate her act in a team. The cooperator ($C$) contributes $c=1$ amount to the team, and the defector ($D$) does not contribute, but free-rides on other $C$ in her team. After this first round interaction, each team member can impose a fine $\beta$ upon each target at a personal cost $\gamma$ for each fine. The punisher ($P$) does this costly behavior against its own benefit. For following discussion, basic parameters are summarized as follows:

\begin{table}[htbp]%
\centering
\begin{threeparttable}
\begin{tabular}{ll}
\toprule
Parameters & Description \\ \midrule
$M$ & The size of total population \\
$G$ & The size of PGG group \\
$r$ & The multiplier of PGG \\
$\beta$ & The amount of punishment on a target per punishment\\
$\gamma$ & The cost of punishment incurred per punishment\\
$\sigma$ & The payoff of loner leaving a team\\
\bottomrule 
\end{tabular}
\end{threeparttable}
\caption{Parameters of PGG. }
\label{TABLE:PGGParam}
\end{table}

For considering stochastic dynamics in finite populations, the groups engaging in a public goods game are given by multivariate hyper-geometric sampling. This sampling affects payoffs of each interaction between two types. Resulting payoffs, fixation probabilities and limiting distribution are given in \aref{APPENDIX:payoff}

\subsection{Stochastic (adaptive) dynamics}

While replicator dynamics provide numerous crucial insights, they are fundamentally based on deterministic dynamics in an arbitrarily large, sometimes infinite, population. Theoretical discussions to overcome this limitation have considered for a long time in various fields such as theoretical ecology, economics or sociology. This paper focuses on a concept developed by economists and evolutionary biologists, stochastic (adaptive) dynamics of finite populations. 

In evolutionary game theory, stochastic (adaptive) dynamics was introduced to understand long-run behavior, which may differ fundamentally from the behavior of the deterministic process by law of large number, replicator dynamics. In replicator dynamics, a state is locally asymptotically stable if any sufficiently small deviation from the original state vanishes. \citet{Young:1993} criticizes this approach because it treats shocks as if they were isolated events. Considering that economic system has constant perturbation from various sources, this assumption of arbitrarily small shock is unsatisfactory. 

Especially, persistent shocks can accumulate and tip the process out of the basin of attraction of asymptotically stable state. Thus, when shock is persistent, generally accepted equilibrium concept, evolutionarily stable strategies, cannot be used to explain long-term behavior of economic system. Especially, this theory can predict the probability of staying in different equilibria independently of the initial conditions. The persistent shocks act as a selection mechanism, and the selection intensity increases the less likely the shocks are. In the long-run distribution relies on the probability of escaping from various states, and this are the function of exponential in error rate. This idea was firstly formalized by \citet{Freidlin_Wentzell:1998}.\footnote{Their idea is that small mutation term makes the system have a different stability for each state, then the limit of invariant distribution can be derived as the mutation probability goes to zero \citep{Ren_Zhang:2008}.} 

Stochastic stability was used to the problem of equilibrium selection in games by \citet{KandoriAND:1993} and \citet{Young:1993}. But, these economic applications are based on ``order-of-magnitude'' comparisons for the transitions between the various recurrent classes of no-mutation process \citep{Ellison:2000}. By this method, one state can be selected as a long-term equilibrium, which is perturbed least by adaptive dynamics, as mutation is trivialized as necessary. 

\citet{Taylor_Fudenberg_Sasaki_Nowak:2005} analyzes a similar but different version of stochastic no-mutation process, where a single mutation can lead to a transition from one absorbing state to another. In this theory, the equilibrium depends on the ``expected speed of flow'' at every absorbing state. This assumes that a single mutant can escape each absorbing state from other types, and the fate of this mutant is determined by fixation probability of two underlying types. Also, \citet{Fudenberg_Imhof:2006} shows that there exists sufficiently small mutation rate that no two individual mutant types cannot coexist. So to speak, the fate of a mutant, its elimination or fixation, is settled before the next mutant appears. Thus the transitions between each homogeneous state occur when a mutant appears and spreads to fixation. 

The advantage of this model is that transition matrix can be nicely formulated by a Markov chain with state space that consists of each homogeneous sate and fixation probability of each state against one another. For this Markov-style transition matrix, unique vectors can be calculated, which is interpreted as invariant distribution of underlying stochastic process. Compared with \citet{KandoriAND:1993} and \citet{Young:1993}, this method shows relative probabilities that each homogeneous state spend with respect to competing others. \aref{APPENDIX:Moran} summarizes the method of \citet{Fudenberg_Imhof:2006}. 

\section{Cooperative vs. Opportunistic Punishment} \label{SEC:Numerical}

To consider different setting and role of punisher in PGG, we look at typical numerical examples by stochastic dynamics. The results in this section are based on \citet{Hauert_Traulsen_Brandt_Nowak_Sigmund:2007} (\aref{APPENDIX:payoff} provides short description of the game and payoff functions). 

\subsection{Sequence of one-shot PGG}

Most of researches including \citet{Hauert_Traulsen_Brandt_Nowak_Sigmund:2007} has assumed an one-shot PGG where players decide strategy or action first, and punish accordingly if some of them want to. In this standard setup, the punisher is originated from $C$ and punishing behavior depends on the information on action choices of players. What if this sequence be reversed? Players commit their punishment first, and choose its action later. For simplicity, commitment is assumed to be always credible, and the number of commitment is announced for participants of a team. For $C$, $D$ and $L$ who do not care for doing punishment, this reversed sequence may not affect their actions. For $P$, however, this information may be crucial in that it conveys information about its own type. Thus, $P$ can choose her action depending on this information. 

By reversing the sequence of PGG, we can discern two types of punisher: cooperative ($CP$) and opportunistic punisher ($OP$). $CP$ commits punishment, but cooperates regardless of information of commitment. $OP$ commits punishment, but chooses whether to cooperate or not depending on the information. If there be few commitment, she might think that $D$ would be better choice to free-ride $C$ or not to be exploited by $D$. We assume that an individual punisher acts cooperatively with probability $q$ that replies on the number of punisher $\tunder{G}{P}$. $q(\cdot)$ is given by 

\begin{align*}
q(\tunder{G}{P}) = \delta \dfrac{ \tunder{G}{P} }{G},
\end{align*}

\noindent where $\delta$ is the responsiveness of punisher. \aref{APPENDIX:payoff} describes payoff functions among $C$, $D$, $L$ and $P$ for $OP$ case. For appropriate parameters, as is discussed in next section, opportunistic punisher makes three different evolutionary dynamics among four types by $\delta$. For the lower range of $\delta$, opportunistic punisher does not make cooperation evolve in a team. Opportunistic punisher makes almost universal cooperation possible for the mid range of $\delta$ when four types can evolved. Finally, for the upper range of $\delta$, $P$ can kill $D$ without the help of $L$. For later discussion, five types of players who are casted in this paper are summarized for their actions and punishing behavior. 

\begin{table}[htbp]%
\centering
\begin{threeparttable}
\begin{tabular}{clc}
\toprule
Name & Action & Punishment \\ \midrule
$C$ & cooperation & No \\ 
$D$ & defection & No \\ 
$L$ & exit & No \\
$CP$\tnote{a} & cooperation & Yes \\ 
$OP$\tnote{b} & conditional cooperation & Yes\\ \bottomrule 
\end{tabular}
{\small
\begin{tablenotes}
\item [a] Cooperative punisher who cooperates unconditionally and punishes defectors in her team.
\item [b] Opportunistic punisher who cooperates depending on the level of punishing commitment.
\end{tablenotes}
}
\end{threeparttable}
\caption{5 types of players}
\label{TABLE:Punisher}
\end{table} 

\subsection{Stochastic dynamics of $CP$ case}

\begin{figure}
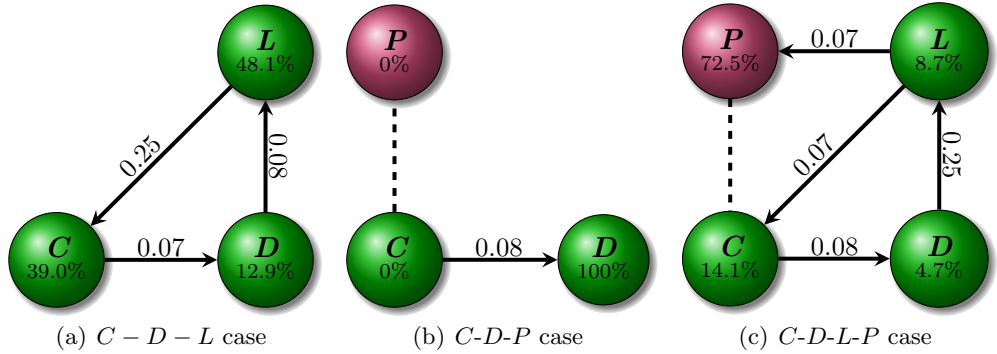

\centering
\subbottom[$C-D-L$ case]{\CDLonesix{0.55}}~
\subbottom[$C$-$D$-$P$ case]{\CDPonesix{0.55}}~
\subbottom[$C$-$D$-$L$-$P$ case]{\CDLPonesix{0.55}}
\caption{$CP$ case stochastic dynamics of PGG. Parameters are given by $s=0.25$, $r=3$, $M=100$, $G=5$, $\beta=1$, $\gamma=0.15$, and $\sigma=1$. The percentage under each type is the relative staying frequency. Arrow from $A$ to $B$ means that $A$ type can fixate $B$ type, and fixation probability is given around the arrow. When a fixation probability is less than $1/M$, it is rare event that one type fixates another. For dashed line, fixation probability is equal to $1/M$ that is the case of random drift. This automata-style presentation helps to illustrate stochastic dynamics of PGG. (a) shows rock-paper-scissor evolutionary cycle among $C-D-L$. (b) is the case that costly punishment cannot survive under $C-D-P$ interaction. Finally, (c) is the evolution of cooperation with four types where $P-C$ random drift consists of around 87{\%}.} 
\label{FIG:HAUERT}
\end{figure}

Main results of \citet{Hauert_Traulsen_Brandt_Nowak_Sigmund:2007} are regenerated in \Fref{FIG:HAUERT}. At first, with voluntary participation, PGG takes circular movement around Cooperator ($C$) $-$ Defector ($D$) $-$ Loner ($L$). The existence of $L$ can perturb universal defection, and make evolutionary cycle for three types, and this can be also observed by replicator dynamics \citep{Hauert_Demonte_Hofbauer_Sigmund:2002}.\footnote{\aref{APPENDIX:PGGRD} provides the technique and results by replicator dynamics.} Even though voluntary participation changes universal defection in PGG, the average payoff a player can get cannot exceed that of $L$. That is, volunteering itself does not enhance the fitness of team members in equilibrium.\footnote{\citet{Sasaki_Okada_Unemi:2007} shows that when players can do mixed strategies of $C$ or $D$ with volunteering, better fitness can be obtained for some parameters.} 

As is shown in \aref{APPENDIX:PGGRD}, assuming infinitely large population, $C-P$ equilibrium can be stabilized for some area in $S_3$ simplex. But, this result just shows that cooperation can be defended only when there already exists sufficient number of punisher. In stochastic dynamic setting, punishment alone cannot police $D$ since the fitness of punisher cannot be higher than that of $D$ ((b) of \fref{FIG:HAUERT}). This can be called ``dilemma of punishment'', which is that $P$ can regulate defective behavior in a group, but the cost of punishment decrease the fitness of $P$. Eventually, unique homogeneous state stochastically stable is $D$ because $P$ cannot be always worse than $D$ in homogeneous state of $D$. In sum, neither of $L$ and $P$ makes any significant contribution to the evolution of cooperation in PGG. 

Interesting dynamics can be made when four types of players are involved in PGG. Panel (c) of \Fref{FIG:HAUERT} shows $C-D-L-P$ interaction. $D$ cannot fixate $P$ because of the existence of $L$. $L$, however, tends to be conquered by $C$ and $P$. The movement between $C$ and $P$ is random drift or neutral selection, where all individuals have the same fitness. For this case, Any random walk in which the probability to move to either side is identical for the transient states leads to the same result. $D$ can be regulated in a circular stochastic relation among four types, and $L$ plays a pivotal role in making a detour for the evolution of cooperation.  \citet{Hauert_Traulsen_Brandt_Nowak_Sigmund:2007} named this mechanism ``via freedom to coercion'', which emphasize synergistic enforcement between $L$ and $P$ in the process. 

The role of $L$ has meaningful economic interpretation where $L$ can be regarded to be the alternative provided by market outside organizations based on human cooperation. Namely, if the group of PGG can be considered as a team or a firm, $L$ represents market. This issue about `Organization vs Market' was treated by \citet{Alchain_Demsetz:1972}. Their conclusion is that monitoring provided by incentive-compatible residual claimant can preserve the comparative advantage of organization over market. Evolutionary dynamics of PGG explore another possibility of regulating issue in team production without formal monitoring or hierarchy. When market provides attractive alternatives, defection in an organization can be regulated in the absence of direct monitoring. In this sense, prolific market and successful organization can co-evolve in our stochastic setting. 

\begin{figure}
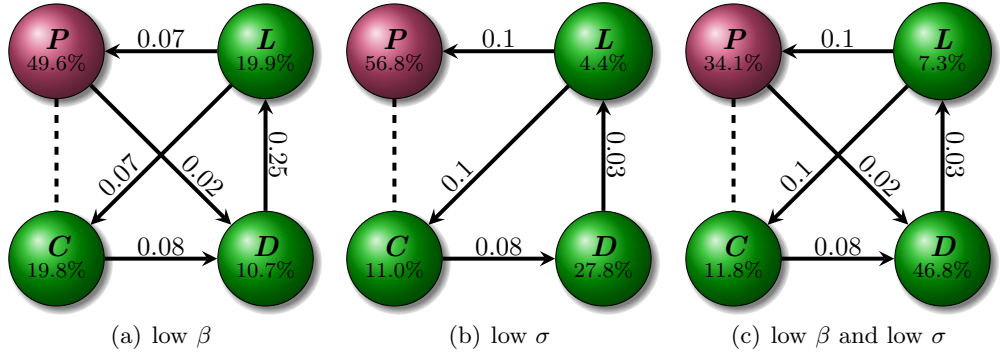

\centering
\subbottom[low $\beta$]{\lowbetaonesix{0.55}}~
\subbottom[low $\sigma$]{\lowsigmaonesix{0.55}}~
\subbottom[low $\beta$ and low $\sigma$]{\lowgammalowsigmaonesix{0.55}}
\caption{$CP$ Stochastic dynamics of PGG when parameters are unfavorable for $CP$. Common parameters are $s=0.25$, $r=3$, $M=100$, $G=5$. Parameters for (a) are $\beta=0.15$, $\gamma=0.15$, $\sigma=1$, those for (b) are $\beta=1$, $\gamma=0.15$, $\sigma=0.1$, those for (c) are $\beta=0.15$, $\gamma=0.07$, $\sigma=0.1$. Low $\beta$ creates $P \to D$ fixation, which decreases the frequency of $P$. Low $\sigma$ makes $D \to L$ fixation slow, and the frequency of $D$ increases. For related calculations, an algorithm is written by Mathematica version 7 of Wolfram Inc.}
\label{FIG:bad16}
\end{figure}

Naturally, the evolution of cooperation in PGG with four types depends critically on underlying parameters. Low $\beta$ activates fixation from $P$ to $D$ ($P \to D$ fixation). \Fref{FIG:bad16} illustrates effects of $\beta$ and $\sigma$. For both parameters, the evolution of cooperation is destroyed as two values decrease. Low $\beta$ creates $P \to D$ fixation, which allows $D$ to absorb both from $C$ and $P$. As $\sigma$ decreases, the flow of $D \to L$ also slows down, which increases staying frequency at $D$-state. (c) of \Fref{FIG:bad16} also shows that lowering $\gamma$ does not change the frequency of cooperative states, staying at $C$ or $P$. 

\subsection{Stochastic dynamics of $OP$ case}

\begin{figure}
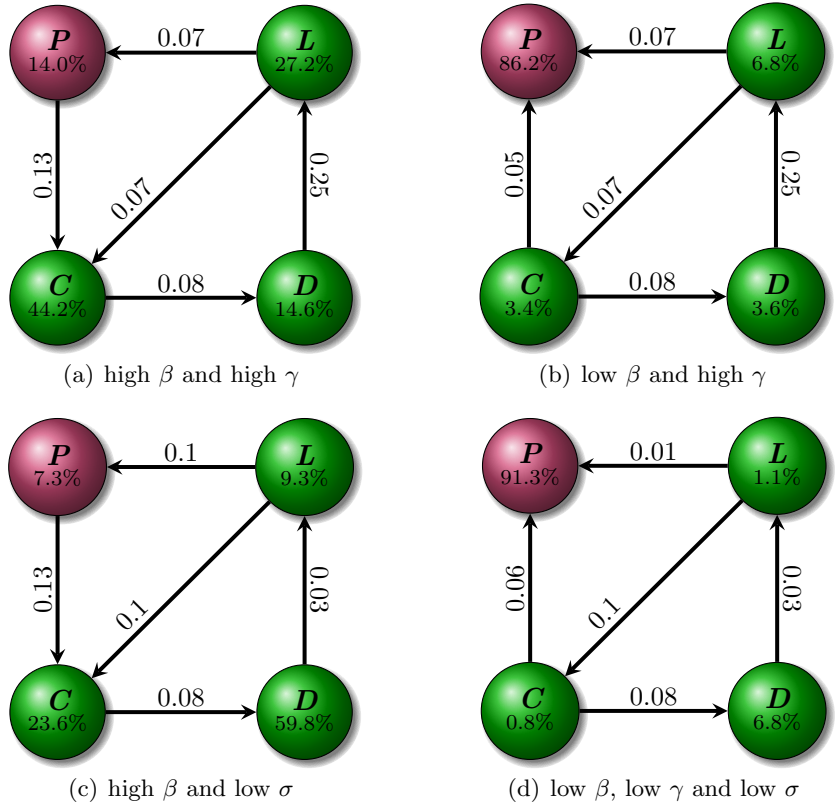

\centering
\subbottom[high $\beta$ and high $\gamma$]{\highbetahighsigmasevensix{0.65}}~~~~~~~~~~
\subbottom[low $\beta$ and high $\gamma$]{\lowbetahighsigmasevensix{0.65}}\\
\subbottom[high $\beta$ and low $\sigma$]{\highbetalowsigmasevensix{0.65}}~~~~~~~~~~
\subbottom[low $\beta$, low $\gamma$ and low $\sigma$]{\lowbetalowgammalowsigmasevensix{0.65}}
\caption{Stochastic dynamics of PGG when $P$ is opportunistic. One-shot PGG proceeds to commit punishment, and choose action. Common parameters are $s=0.25$, $r=3$, $M=100$, $G=5$, $\delta=0.8$. Parameters for (a) are $\beta=1$, $\gamma=0.15$, $\sigma=1$, those for (b) are $\beta=0.15$, $\gamma=0.15$, $\sigma=1$, those for (c) are $\beta=1$, $\gamma=0.15$, $\sigma=0.1$, those for (d) are $\beta=0.15$, $\gamma=0.07$, $\sigma=0.1$. As the figure shows, $P \to C$ fixation is key to the evolution of cooperation. This dynamics is made by opportunism by $P$, which decreases the frequency of $C$, and exploitation by $D$ is prevented.}
\label{FIG:good76}
\end{figure}

As is stated, players declare their commitment on punishing first, and choose actions in $OP$ case. Assuming independence in choosing actions, the expected number of $P$ in a team, which is equal to the number of commitment, is simply given by $q \cdot \tunder{G}{P}$. payoffs are modified for opportunistic punisher, and stochastic dynamics of $OP$ case is given in \Fref{FIG:good76}. 

Numerical examples implies that $OP$ can contribute the evolution of cooperation in which $CP$ loses her power as long as $\delta$, the responsiveness of punisher, is sufficiently high. They imply that the efficacy of $OP$ comes from exploiting $C$ and fighting $D$ more successfully. When $\beta$ and $\gamma$ is sufficiently high, $OP$ cannot contribute anymore because the commitment of punishing hurts herself to a serious level. Thus, when punishing is more effective than a certain level, $CP$ is more effective than $OP$ in fostering the evolution of cooperation in a PGG team. 

\begin{figure}
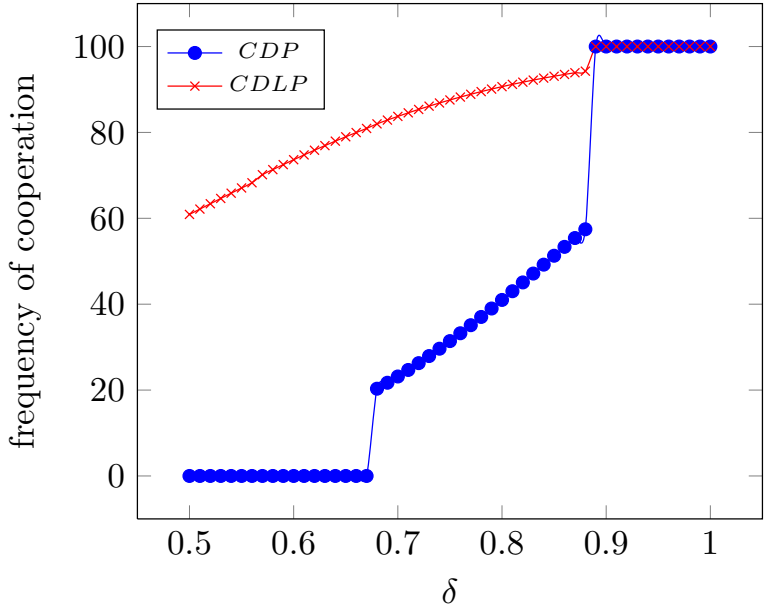
\centering
\deltawork{1.2}
\caption{The evolution of cooperation by $OP$ in $C-D-P$ and $C-D-L-P$. Parameters are equal to those in (d) of \Fref{FIG:good76}. $Y$-axis is sum of frequency in $C$ and $P$. For $\delta < 0.88$, only four types, $C-D-L-P$ can make cooperation evolve. As $\delta$ grows, the evolution of cooperation can be made only by $C-D-P$ interaction. For this case, $L$ as a depressor of $D$ is unnecessary.}
\label{FIG:ThreevsFour}
\end{figure}

\Fref{FIG:ThreevsFour} shows an interesting dynamics of $OP$ case. As $\delta$ approaches $1$, for some proper parameters, the help of the loner can be redundant. $C-D-OP$ dynamics make $100{\%}$ state of cooperation at $P$. When $\delta$ is sufficiently high, $P$'s selfishness alone makes the evolution of cooperation. 

\begin{figure}
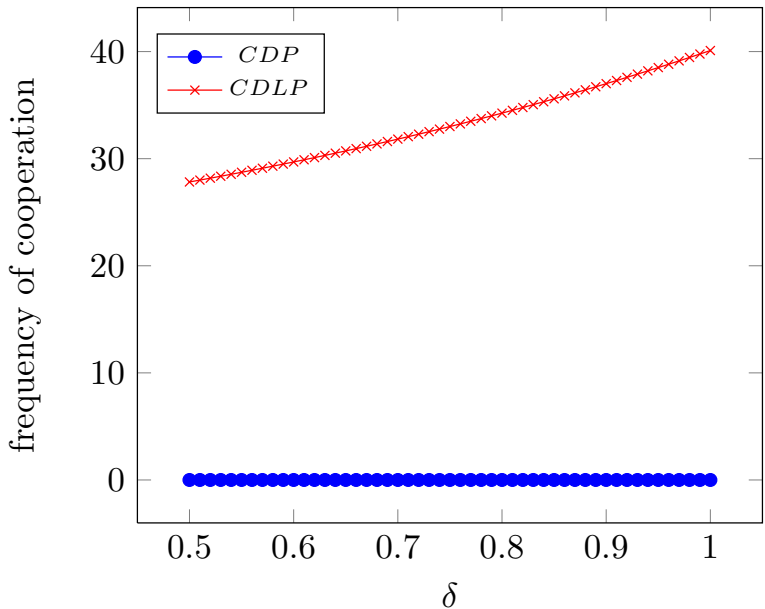
\centering
\deltaerror{1.2}
\caption{The evolution of cooperation when $P$ plays $C$ with error or tremble rate $\delta$. All parameters are equal to \Fref{FIG:ThreevsFour}. When $\delta=0~(1)$, $P$ plays $D~(C)$ always. Results imply that simple error cannot increase cooperation in a team compared to $CP$. This shows the strategical advantage of $OP$ who plays more sophisticated strategy than simple types of error. }
\label{FIG:ErrorP}
\end{figure}

Let us compare stochastic dynamics of PGG to replicator dynamics. \aref{APPENDIX:PGGRD} illustrates replicator dynamics of PGG. For $C-D-OP$ interaction, equilibria by replicator dynamics agree well with equilibrium by stochastic dynamics. When parameters are proper, $C-D-OP$ interaction by replicator dynamics make two type of NE . For this case, when $\beta$ is low (but not too much), $\gamma$ is low, and $\delta$ is high, $C-OP$ mixture with high density of $OP$ is NE. This is a equilibrium state that opportunism by $P$ is sparsely observed because of high frequency of $P$. As is discussed, stochastic dynamics select this almost full cooperative state by $OP$ over $D$-state when parameters are proper. 

Now, we consider how the opportunism of $P$ helps the evolution of cooperation. Let us compare the behavior of $OP$ to simple tremble or error in playing action. As \Fref{FIG:ErrorP} shows simple tremble does not help to overcome invasion of $D$ or fixating $C$, which is key part that $OP$ plays. This implies that opportunistic punisher has more sophisticated strategic reaction than simple types of error. 

The opportunism makes $P$ play in a {\itshape correlated} way according to the composition of her team informed by the level of commitment. For example, when a team consists all of $C$ or $D$ except a $P$, $OP$ hardly do $C$ because commitment level is low, and punishment from others is not expected. Thus, for $C$-only team case, the payoff of $OP$ is higher than $CP$. For $D$-only team case, the payoff of $OP$ is less than that of $D$, but higher than that of $CP$. In the opposite instance, if a team consists all of $P$ except one $P$, $OP$ always does $C$ because playing $D$ hurts her by ample of punishment. For this case $P$ resembles $CP$. That is, assuming proper size of $G$ and sets of parameter such as $\beta$, $\gamma$ and $\delta$, payoff of $OP$ in three pure state are given by

{\small
\begin{align*}
\hspace{-0.8cm}
\begin{cases}
\tunder{\pi}{OP} (=\frac{G-1}{G} r) > \tunder{\pi}{CP} (=\tunder{\pi}{C}=r-1)~&\text{for all-$C$ case} \\
\tunder{\pi}{D} (=0-\beta) \approx \tunder{\pi}{OP}(=0-(G-1)\gamma) > \tunder{\pi}{CP} (=-1-(G-1)\gamma)~&\text{for all-$D$ case} \\
\tunder{\pi}{OP} \approx \tunder{\pi}{CP} (=r-1)~&\text{for all-$P$ case},
\end{cases}
\end{align*}
}

\noindent where $\tunder{\pi}{k}$ is the payoff of type $k$. 

This strategic flexibility comes from nonlinearity made by probabilistic reaction modeled by $q(G_p)$. When the degree of nonlinearity, $\delta$, is sufficiently high, $OP$ can copy better reaction between $C$ and $D$ in correlated way. This flexibility creates $C \to P$ fixation, and ends $P \to D$ fixation. 

\section{Calculating Fixation Probabilities by Fermi Function}\label{SEC:Fermi}

As \aref{APPENDIX:payoff} shows, fixation probabilities of Moran process can be defined within a certain boundary of $s$, the intensity of selection. To generalize our model, a pair-wise comparison by Fermi function is to be introduced \citep{Traulsen_Nowak_Pacheco:2006, Altrock_Traulsen:2009}. Fermi function defines dynamics of payoff difference between two types for any $s$.\footnote{If we simply replace the fitness function of Moran process, $(1-s)+s \pi$, with $e^{(1-s)+s \pi}$ to considering any $s \in [0,1]$, resulting term for fixation probability is identical.} When transition from $A$-type to $B$-type occurs, the probability is assumed to be 

\begin{align*}
p=\dfrac{1}{1+e^{s (\pi_A - \pi_B)}},
\end{align*}
\noindent which is called called Fermi function. This makes
\begin{align*}
\dfrac{T^-_j}{T^+_j} = e^{-s(\pi_A - \pi_B)}.
\end{align*}

The evolution of cooperation can be analyzed by investigating fixation probabilities between four types. Specifically, as is implied in numerical examples in \Sref{SEC:Numerical}, the evolution of cooperation may depend on the fixation between $P$ and $D$, and that between $P$ and $C$. 

\begin{figure}
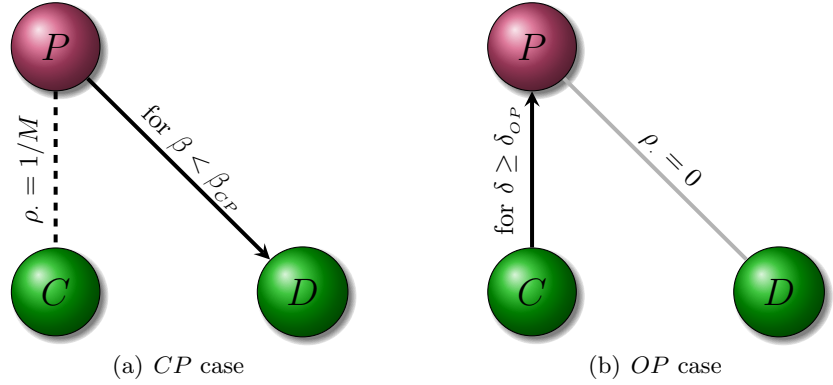

\centering
\subbottom[$CP$ case]{\CDPFermiCP{0.65}}~~~~~~~~~~~
\subbottom[$OP$ case]{\CDPFermiOP{0.65}} \\
\caption{Key fixation diagrams of $CP$ and $OP$. We assume proper parameters for unrelated variables. When $\beta$ is larger than a critical level, $D$ can fixate $CP$ and the evolution of cooperation is made. When $CP$ is less effective, $OP$ can absorb $C$ as long as $\delta$ is higher than a certain level. This prevents $D$ from exploiting $C$, and the cooperation can be evolved by players' opportunistic behavior.} 
\label{FIG:Fermi}
\end{figure}

We apply three approximations to get analytic expression. 
\begin{inparaenum}[1)]
\item As $1/M \to 0$ can be assumed for sufficiently high $M$, related payoffs can be linearized around $1/M \approx 0$ as many as necessary.
\item Approximated fixation probabilities are categorized into two; the one is surely larger than $1/M$, the other cannot exceed it. We take the first kind as legitimate, and set the second to be $0$.
\item $\sum_{k=1}^{k} \cdots \approx \int_{1}^{k} \cdots d k$ can be used because the error between two is $O((\frac{1}{M})^2)$, which is plausible for fairly larger $M$.
\end{inparaenum}

Let $\rho_{ij}$ denote this simplified fixation probability of single $i$-type in the population that consists totally of $j$-type. Common fixation probabilities for each type of punisher are given by 

\begin{align*}
\tunder{\rho}{DC}&=s(1-\dfrac{r}{G}) \\
\tunder{\rho}{LD}&=s\sigma \\
\tunder{\rho}{CL}&=\dfrac{2\sqrt{2s(G-1)(r-\sigma-1)}}{2\pi +2\sqrt{2s(G-1)(r-\sigma-1)}}.
\end{align*}

At first, these are invariant by the type of punisher. High intensity increases $\tunder{\rho}{DC}$ and $\tunder{\rho}{LD}$. When $\sigma$ increases(decreases), the flow of $D \to L$ speeds up(down), but that of $L \to C$ speeds down(up). As is discussed in \Sref{SEC:Numerical}, when $\sigma$ gets smaller, the frequency of $L$ decreases, and that of $C$ increases consequently. Without $P$, this flow ends up with higher frequency of $D$ due to lower $\tunder{\rho}{LD}$. 

At first, \Fref{FIG:Fermi} shows $P \to D$ and $P \to C$ fixation diagrams for each type of punisher. When $\beta < \tunder{\beta}{CP}$ where where $\tunder{\beta}{CP} \equiv \frac{G-r}{G(G-1)}$, fixation probabilities are respectively given by 

\begin{align*}
\begin{cases}
\tunder{\rho}{DP}=s [1-\dfrac{r}{G}-(G-1)\beta] & \text{for $CP$} \\
\tunder{\rho}{DP}=0 & \text{for $OP$} 
\end{cases}
\end{align*}

(a) of \Fref{FIG:Fermi} shows when $CP$ is effective. When $\beta < \tunder{\beta}{CP}$, $D$ can fixate $P$. The evolution of cooperation is hindered as numerical examples shows. For this instance, $OP$ who plays opportunistically can make cooperation evolve in a team. $C \to P$ fixation is a key mechanism, which weakens $C \to D$ fixation. Different from $CP$ case where $P$ and $C$ have equal fitness, $OP$ has higher fitness than $C$ because $OP$ tends to play $D$ more as there exists more $C$. As the frequency of $C$ decreases due to $C \to P$ fixation, the relative staying at $D$ does also. Numerical examples shows that $\tunder{\rho}{PC}$ for $OP$ changes discontinuously for a critical level of $\delta$, $\tunder{\delta}{OP}$. This implies that players' responsiveness to the information can have a pivotal role in fostering the evolution of cooperation. 

The analytic approach for this intuition can be done by Fermi function. For $CP$, $\tunder{\rho}{PC}$ is given by $1/M$. For $OP$, when $\beta$ and $\gamma$ are low as necessary, 

$\tunder{\rho}{PC}$ is calculated as
\begin{align*}
\begin{cases}
\tunder{\rho}{PC}=0 & \text{for $\delta < \tunder{\delta}{OP}$} \\
\tunder{\rho}{PC}= \frac{1}{\frac{G^2 \exp[-\frac{s \left(G^2-G (r+1)+r \delta \right)}{G^2}]}{G s
(G-r-1)+r s \delta }+1}& \text{for $\delta \geq \tunder{\delta}{OP}$},
\end{cases}
\end{align*}

\noindent where $\tunder{\delta}{OP} \equiv \frac{G \left(G^2 (\beta +\gamma )-G (\beta +\gamma +1)+2 r+1\right)}{(2 G-1) r+2(G-1) G (\beta +\gamma )}$. By Fermi function, the fixation probability of $\tunder{\rho}{PC}$ for $OP$ behaves nicely in discontinuous way as numerical examples do. 

\section{Concluding Remarks}\label{SEC:Concluding}

For the gaming situations in which Nash equilibrium predicts general defection, the possibility of cooperation is one of the most challenging and crucial questions of evolutionary economics and biology. This paper, in stochastic dynamic setting, discusses an intriguing and paradoxical path to cooperation via players' opportunistic behavior. Different from \citet{Hauert_Traulsen_Brandt_Nowak_Sigmund:2007} that emphasizes the role of quitting to support $P$ who can regulate $D$, by reversing the sequence of PGG, we propose that the opportunistic behavior of $P$ may paradoxically make cooperation evolve in a team. Moreover, for the cases that altruistic punisher cannot help the evolution of cooperation, our opportunistic punisher can. This comes from the dual role of opportunism: $OP$ can end $P \to D$ fixation, and make $C \to P$ fixation. Both fixating flows decrease relative staying frequency at $D$-state, which encourages the evolution of cooperation. 

Finally, two future research agenda is to be mentioned. First, in this paper, we regard $\delta$ as parameter, which determines the responsiveness of probabilistic opportunism. Even though simplicity justifies this, more interesting results and questions can be discussed if we make $\delta$ determined endogenously. Also, another simplification is that the commitment of punishing is always credible. In real world, some forms of contracts are done in this fashion by depositing some of money to a third party for the case of non-fulfillment. However, partial credibility of commitment may reveal more interesting and unexpected results on the issues of this paper. 

\appendix 
\appendix\def\thesection{Appendix \Alph{section}}
\setcounter{equation}{0}
\numberwithin{equation}{section}
\def\theequation{\Alph{section}.\arabic{equation}}

\section{The Stochastic Dynamics of Generalized Moran Process}\label{APPENDIX:Moran}
\subsection{Moran process}
Moran process is a classical model of population that is developed in population genetics, and has been imported to game theory recently. In every time step an individual is randomly chosen for reproduction by its fitness, and makes a single clone that replace a randomly selected other member. Moran process represents a simple birth-death process. For the whole process, the size of total population, $M$, remains constant, i.e., Moran process ignores effects of population size. This assumption of exogenous finite population size can be considered as an approximation to a model where environmental forces keep the population from becoming infinite \citep{Fudenberg_Imhof_Nowak_Taylor:2004}. 

For studying finite populations, it is convenient to transform fitness into convex combination of baseline fitness (generally assumed to be $1$) and payoff obtained from interaction. That is, $f=(1-s)1+s \pi$ where $f$ is fitness of a player, $\pi$ is the payoff from the game. $s$ controls the intensity of selection. When $s=0$, selection is neutral and we have random drift. For $s \to 1$, fitness can be equated to payoff. Since $f$ should be positive, there exists maximum $s$. 

\subsection{Fixation probability}

Repeatedly applying Moran updating determines the evolutionary result of residents and mutants. In the absence of mutations, which is in the spirit of literature on large deviations of long-run behavior, Moran process ends up with a homogeneous population with all residents or all invaders \citep{Young_Foster:1990,KandoriAND:1993,Young:1993,Kandori_Rob:1995}. Regardless of initial state of population, eventually all members of the population consists of one type. When this homogeneous state by one type is realized, conquering type is said to reach {\itshape fixation}. It is the key to this dynamics to find fixation probabilities of types in the population. 

Let us explain how to find fixation probabilities by two-strategies case. For $M$-size population, the number of $A$-strategy players is $j$, and the number of $B$ is $M-j$. The probability to increase the number of $A$ from $j$ to $j+1$ is denoted by $T^{+}_j$. Similarly, $T^{-}_j$ is probability to decrease $j$ by $1$. Considering that there exist two absorbing states with no-mutation game dynamics, two fixation probabilities is given by 

\begin{align*}
\phi_0 = 0~\text{and}~\phi_M=1
\end{align*}

\noindent where $\phi_j$ is the fixation probability where the number of $A$ is $j$. For intermediate state, the fixation probability are given by 

\begin{align} \label{EQ:fixation:02}
\phi_j = T^{-}_j \phi_{j-1} + (1-T^{-}_j-T^{+}_j)\phi_j + T^{+}_j \phi_{j+1},
\end{align}
\noindent which is an expression of fixation probability by its one back-and-forth time step. Rearrange \Meref{EQ:fixation:02} makes 
\begin{align} \label{EQ:fixation:02b}
0 =- T^{-}_j (\phi_{j} -\phi_{j-1} ) + T^{+}_j ( \phi_{j+1} - \phi_{j} ).
\end{align}

\Meref{EQ:fixation:02b} can be suitably used to make a recursion for the differences between fixation probabilities. For our discussion $\phi_1$, the fixation probability of a single $A$ individual, is particularly important. By some algebra, this is calculated as 

\begin{align}\label{EQ:fixation:03}
\phi_1=\dfrac{1}{\displaystyle\sum_{k=1}^{M-1} \prod_{j=0}^{k} \frac{T^-_j}{T^+_j}}.
\end{align}

It is possible to calculate fixation probability for any initial state of existing $i$-number of $A$, $\phi_i$ \citep{Nowak_Sasaki_Taylor_Fudenberg:2004b,Taylor_Fudenberg_Sasaki_Nowak:2005}. Only $\phi_1$ is needed to investigate stationary distribution with small mutations. 

For neutral selection where drift is purely random, $T^-_j = T^+_j$ holds, hence $\phi_1$ is easily given by $1/M$. This fixation probability of random drift is used to judge how strong a single individual enough to fixate whole population. When the fixation probability of a specific individual of a type is larger than $1/M$, there is a statistical tendency for this type to occupy the whole population. Otherwise, this type is easy to be fixated by other types whose fixation probabilities are larger than $1/M$. This criteria about fixation has a good interpretation to describe mutual invasion between two types, which is useful for our purpose.\footnote{When fixation probability from $A$ to $B$ is smaller than $1/M$, we can ignore this direction of movement. This qualitative approach makes analysis simpler and illustrative as following automata-style diagram shows}. 

\section{Payoffs of PGG}\label{APPENDIX:payoff}

We denote the number of cooperator by $c$, defector by $d$, loner by $l$, and punisher by $p$. Naturally, $M=c+d+l+p$ holds. Also, $0<\sigma+1<r<G$, and $G \geq 3$ are assumed for relevant discussion. $\tunder{\pi}{CD}$, the expected average payoff of focal $C$ against $D$, is given by 

\begin{align*} 
\tunder{\pi}{CD}=\sum_{k=0}^{G-1} \underbrace{\dfrac{\binom{c-1}{k} \binom{M-c}{G-k}}{\binom{M-1}{G-1}}}_{\text{(i)}} \underbrace{\vphantom{\dfrac{\binom{c}{k}}{\binom{M}{G}}}\left( \dfrac{k+1}{G} r -1 \right)}_{\text{(ii)}},
\end{align*}

\noindent where 

\begin{inparaenum}[(i)]
\item is the probability that there are $k$ number of $C$ and $G-k$ number of $D$ when $G$-sized team is made from $M$-population,
\item is the average payoff from $k$-number $C$ team.
\end{inparaenum} 
Relevant payoffs for $D$ and $L$ are given by 

\begin{align*} 
\tunder{\pi}{DC}&=\sum_{k=0}^{G-1} \dfrac{ \binom{c}{k}\binom {M-c}{G-k} }{ \binom{M-1}{G-1} } \left( \dfrac{k}{G} r \right) \\
\tunder{\pi}{LC}&= \pi_{LD} = \sigma \\
\tunder{\pi}{CL}&=\left[ 1-\dfrac{\binom{l}{G-1}} {\binom{M-1}{G-1}} \right] (r-1) + \dfrac{\binom{l}{G-1}} {\binom{M-1}{G-1}} \sigma.
\end{align*}

Interactions of $C$, $D$ and $L$ with respect to $P$ are specified by the type of $P$. When punisher is cooperative type, $CP$, related payoff are 

\begin{align*} 
\tunder{\pi}{CP} &= \pi_{PC} = r-1 \\
\tunder{\pi}{DP} &= \sum_{k=0}^{G-1} \dfrac{\binom{d-1}{k} \binom{M-d}{G-k-1}}{\binom{M-1}{G-1}} \left[ \dfrac{G-k-1}{G} r - (G - k - 1) \beta \right] \\
\tunder{\pi}{PD} &= \sum_{k=0}^{G-1} \dfrac{\binom{p-1}{k} \binom{M-p}{G-k-1}}{\binom{M-1}{G-1}} \left[ \dfrac{k+1}{G} r -1- (G - k - 1) \gamma \right] \\
\tunder{\pi}{LP} &= \sigma \\
\tunder{\pi}{PL} &= \dfrac{\binom{M-p}{G-1}} {\binom{M-1}{G-1}} \sigma +\sum_{k=1}^{G-1} \dfrac{\binom{p-1}{k} \binom{M-p}{G-k-1}}{\binom{M-1}{G-1}} (r-1).
\end{align*}

The payoffs for $OP$ are 

{\tiny
\begin{align*} 
\tunder{\pi}{CP} &= \sum_{k=0}^{G-1} \dfrac{\binom{c-1}{k} \binom{M-c}{G-k-1}}{\binom{M-1}{G-1}} \left[ \dfrac{(k+1)+\delta\frac{G-k-1}{G}(G-k-1)}{G} r - 1 \right] \\
\tunder{\pi}{PC} &= \sum_{k=0}^{G-1} \dfrac{\binom{p-1}{k} \binom{M-p}{G-k-1}}{\binom{M-1}{G-1}} \left[ \dfrac{(G-k-1)+\delta\frac{k+1}{G}(k+1)}{G} r -\frac{k+1}{G}-(1 - \delta\frac{k + 1}{G} k)(\beta+\gamma) \right] \\
\tunder{\pi}{DP} &= \sum_{k=0}^{G-1} \dfrac{\binom{d-1}{k} \binom{M-d}{G-k-1}}{\binom{M-1}{G-1}} \left[ \dfrac{(G-k-1)\delta \frac{G-k-1}{G}}{G} - (G - k - 1) \beta \right] \\
\tunder{\pi}{PD} &= \sum_{k=0}^{G-1} \dfrac{\binom{p-1}{k} \binom{M-p}{G-k-1}}{\binom{M-1}{G-1}} \left[ \dfrac{(k+1)\delta\frac{k+1}{G}}{G} -\delta\dfrac{k+1}{G}-\left(G-k-1+(1-\delta\dfrac{k+1}{G})k\right)\gamma -(1-\delta\dfrac{k+1}{G})k \beta \right] \\
\tunder{\pi}{LP} &= \sigma \\
\tunder{\pi}{PL} &= \dfrac{\binom{M-p}{G-1}} {\binom{M-1}{G-1}} \sigma +\sum_{k=1}^{G-1} \dfrac{\binom{p-1}{k} \binom{M-p}{G-k-1}}{\binom{M-1}{G-1}} (r-1).
\end{align*}
}

For Moran process, transition probability for one forward step is given by 

\begin{align*}
T^+_{ij}=\underbrace{\dfrac { m_i [(1-s)+s \pi_{ij}]} {m_i [(1-s)+s \pi_{ij}]+(M-m_i)[(1-s)+s \pi_{ij}]}}_{\text{probability for $i$'s reproduction}} \underbrace{\dfrac{M-m_i}{M},}_{\text{probability for type $j$'s death}}
\end{align*}
\noindent where $m_i$ is the number of type $i$, and $i,j \in \{C,D,L,P \}$. the probability for one backward step is given by 
\begin{align*}
T^-_{ij}=\dfrac{(M-m_i)[(1-s)+s \pi_{ij}]} {m_i [(1-s)+s \pi_{ij}]+(M-m_i)[(1-s)+s \pi_{ij}]} \dfrac{m_i}{M}.
\end{align*}

Fixation probability for $i$ against $j$ is 

\begin{align*}
\phi_{ij}=\dfrac{1}{\displaystyle \sum_{k=0}^{M-1} \prod_{m_1=1}^k \frac{T^-_{ij}}{T^+_{ij}}}=\frac{1}{\displaystyle \sum_{k=0}^{M-1} \prod_{m_1=1}^k \frac{1-s + s \pi_{ji}}{1-s+s\pi_{ij}}}.
\end{align*}

As the fitness should be positive for proper $\phi_{ij}$, an upper limit on $s$ is given by $1/(1-\min {\pi_{ij}})$. 

Fixation probabilities are used for making a Markov transition matrix between four different homogeneous states, which is 

{\tiny
\begin{align*}
\hspace{-1cm}\left( 
\begin{array}{cccc}
1-\tunder{\phi}{DC} -\tunder{\phi}{LC} - \tunder{\phi}{PC} & \tunder{\phi}{CD} & \tunder{\phi}{CL} & \tunder{\phi}{CP} \\
\tunder{\phi}{DC} & 1-\tunder{\phi}{CD} -\tunder{\phi}{LD}-\tunder{\phi}{PD} & \tunder{\phi}{DL} & \tunder{\phi}{DP} \\
\tunder{\phi}{LC} & \tunder{\phi}{LD} & 1-\tunder{\phi}{CL}-\tunder{\phi}{DL}-\tunder{\phi}{PL} & \tunder{\phi}{LP} \\
\tunder{\phi}{PC}& \tunder{\phi}{PD} & \tunder{\phi}{PL} & 1-\tunder{\phi}{CP}-\tunder{\phi}{DP}-\tunder{\phi}{LP} 
\end{array}
\right).
\end{align*}
}

Above matrix defines entry and exit between four homogeneous states. For example, first row describes how one mutant $C$ influences system. The second elements of this row shows the probability that a single mutant $C$ conquers or fixates $D$-homogeneous state. Naturally, sum of each column should be 1. \citet{Fudenberg_Imhof:2006} shows that normalized eigenvector to the biggest eigenvalue, $1$ for this case, determines stationary distribution for small mutations. 

\section{Replicator Dynamics and Stable NE} \label{APPENDIX:PGGRD}

\begin{figure}[h]
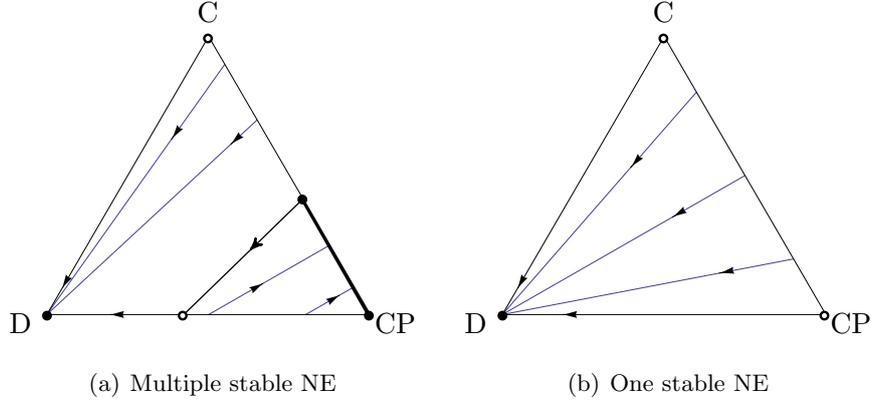

\centering
\subbottom[Multiple stable NE]{\CDAPGOOD{0.35}}
\subbottom[One stable NE]{\CDAPBAD{0.35}}
\caption{Solid circle represents stable NE, and empty circle does unstable fix points. (a) shows that $C-CP$ co-existence can be stable equilibria for some area of $S_3$ simplex. (b) shows that this evolution of cooperation disappears when underlying conditions turn severe. Figures are generated by the modified version of DYNAMO originally written by William Sandholm, Emin Dokumaci, and Francisco Franchetti. (\url{http://www.ssc.wisc.edu/~whs/dynamo/index.html})}
\label{FIG:APPENDIX:CDAP}
\end{figure}

Replicator dynamics for PGG in the paper can be derived by formulating payoff for each type of player. For $C-D-CP$, 

\begin{align*}
\tunder{\pi}{C} & = r (c+p) - 1 \\
\tunder{\pi}{D} & = r (c+p)-\beta p (G-1) \\
\tunder{\pi}{P} & = r (c+p)-1-\gamma d (G-1)
\end{align*}

\noindent where $c$, $d$ and $p$ denotes relative frequency of $C$, $D$ and $P$ in a infinitely large population respectively. This system can be treated by a system of linear differential equations, and phase diagram and stable NE can be easily given. Phase diagrams of \Fref{FIG:APPENDIX:CDAP} shows two types of equilibrium state. For $\beta > 1/(G-1)$, multiple stable NE are obtained as (a) of \Fref{FIG:APPENDIX:CDAP}. That is, when punishment are sufficiently effective, the continuum of $C-CP$ mixture can be supported as stable NE. For $\beta \leq 1/(G-1)$, $D$-state is unique NE. Without exit option, stochastic dynamics selects $D$-state as unique equilibrium for $C-D-CP$. 

\begin{figure}[h]
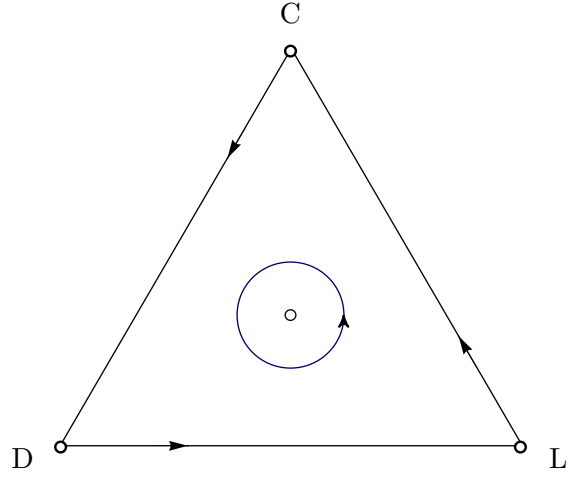

\centering
\CDLCYCLE{0.5}
\caption{$C-D-L$ interaction shows Rock-paper-scissor dynamics.}
\label{FIG:APPENDIX:CDL}
\end{figure}

For $C-D-L$ case, payoffs are given by 

\begin{align*}
\tunder{\pi}{C} & = (1-l^{G-1})(r c - 1) + l^{G-1} \sigma \\
\tunder{\pi}{D} & = (1-l^{G-1})(r c)+ l^{G-1} \sigma \\
\tunder{\pi}{L} & =\sigma,
\end{align*}

\noindent where $l$ denotes the relative frequency of $L$. Different from $C-D-CP$ case, $C-D-L$ cannot be easily treated because $L$ makes the system nonlinear one. \citet{Brandt_Hauert_Sigmund:2003} gives a trick to formulate replicator dynamics. This makes use of the fact that payoff difference between $C$ and $D$ depends only on $l$. Three homogeneous states are natural fixed points. There are no other fixed points on the boundary of $S^3$ simplex. For $r>2$, unique rest point in interior of $S^3$, and interior dynamics can be described by Hamiltonian system. This is equivalent to rock-paper-scissor dynamics where rest point is surrounded by periodic orbits as is shown in \Fref{FIG:APPENDIX:CDL}. 

\begin{figure}[h!]
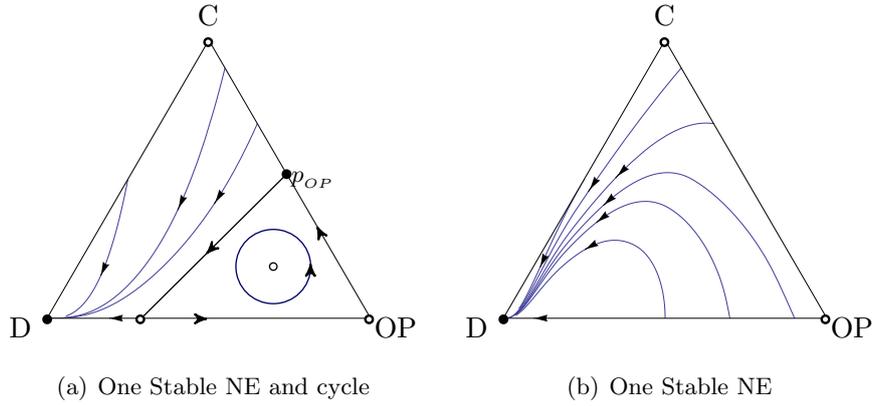

\centering
\subbottom[One Stable NE and cycle]{\CDOPBAD{0.35}}
\subbottom[One Stable NE]{\CDOPGOOD{0.35}}
\caption{(a) shows the case for two stable NE, which are $D$-state and $C-OP$ coexistence. (b) shows the case for one stable NE, $D$-state.}
\label{FIG:APPENDIX:CDOP}
\end{figure}

Finally, payoff for $OP$ case are given by 

\begin{align*}
\tunder{\pi}{C} & = r (c+(\delta p) p) - 1 \\
\tunder{\pi}{D} & = r (c+(\delta p) p)-\beta p (G-1) \\
\tunder{\pi}{P} & = r (c+(\delta p) p)-\delta p - \gamma (d +(1-\delta p) p )(G-1)-\beta (1-\delta p) p (G-1)
\end{align*}

By using similar method of $C-D-L$ case, it can be checked that there is no fixed point in interior of $S_3$ simplex, and unique NE exist at $D$ and/or $C-OP$ boundary.\footnote{To find dynamic path, payoff differences between $C$ and $OP$, that between $D$ and $OP$ should be checked.} \fref{FIG:APPENDIX:CDOP} shows two typical dynamics for $C-D-OP$ case. For $\gamma>0$, the condition is given by 

\begin{align*}
\begin{cases}
\text{(a)}&~\text{for}~\beta > \dfrac{1}{G-1},~\dfrac{-\gamma+G \gamma}{-1-\beta+G \beta -\gamma + G \gamma} < \delta <1 \\
\text{(b)}&~\text{for otherwise}.
\end{cases}
\end{align*}

Cooperative equilibrium made by $OP$ in stochastic dynamics is the case of (a) with low $\beta$, low $\gamma$ and high $\delta$. $\tunder{p}{OP}$  in (a) of \Fref{FIG:APPENDIX:CDOP} is given by $\frac{1}{(\beta+\gamma)(G-1)}$. High $\beta$ and $\gamma$ make $\tunder{p}{OP}$ small, which is that region for evolutionary cycle is enlarged. Otherwise, when region for evolutionary cycle shrinks, population consists mostly of $C$ and $OP$, which can be regarded as cooperative state. 

\bibliography{3rdDraft_arXig_Total}
\end{document}